\documentclass[twocolumn]{jpsj2} 
\def\simge{\lower0.7ex\hbox{$\ \overset{>}{\sim}\ $}}
\def\simle{\lower0.7ex\hbox{$\ \overset{<}{\sim}\ $}}
\usepackage{color}
\usepackage{multirow}
\title{Antiferro Quadrupole Orders in Non-Kramers Doublet Systems}

\author{Kazumasa HATTORI and Hirokazu TSUNETSUGU}

\inst{Institute for Solid State Physics, University of Tokyo, 5-1-5, Kashiwanoha, Kashiwa, Chiba 277-8581, Japan
}

\author{Kazumasa HATTORI$^{1,2}$\thanks{E-mail:
hattori@issp.u-tokyo.ac.jp} and Hirokazu TSUNETSUGU$^1$}
\inst{$^1$Institute for Solid State Physics, University of Tokyo, 5-1-5, Kashiwanoha, Kashiwa, Chiba 277-8581, Japan \\
$^2$Institut f\"{u}r Theoretische Physik, Universit\"{a}t zu K\"{o}ln,
Z\"{u}lpicher Str., 77, D-50937 K\"{o}ln, Germany}

\abst{
We investigate antiferro quadrupole orders in systems with non-Kramers doublet
ground state with total angular momentum $J=4$ in $T_d$ point
group symmetry. We demonstrate that a pure $O_2^2$
antiferro quadrupole order is impossible  in general crystalline-electric field
potential and should be accompanied by 
ferro $O_2^0$ quadrupole moment. The temperature and magnetic-field
phase diagram is obtained by mean-field approximation of intersite quadrupole
interactions and the excitation spectrum is analyzed by ``spin''-wave approximations.
Gapless excitations emerge at the boader of antiferro $O_2^2$ 
quadrupole phases under magnetic field. Quadrupole susceptibilities in
the antiferro quadrupole ordered state exhibit unusual singularity and
especially the uniform quadrupole susceptibility diverges in addition to
the staggered ones. These unusual
singularities are also realized at the critical field along [111] direction. 
We also discuss recent
experimental results in Pr$T$$_2X_{20}$($T$=Ir, Rh, Ti, V, and $X$=Zn, Al).
 }

\begin{document}
\maketitle

\section{Introduction}
Orbital order appears in varieties
of systems in condensed matter physics such as in $d$- and $f$-electron
strongly correlated systems with orbital degrees of freedom.\cite{Tokura,Kura} 
Quadrupole orders are typical and most intensively studied cases. 
Such
orbital orders can be in principle described by similar theoretical
approaches as in spin systems. As is evident from the nature of orbital
degrees of freedom, interactions have spatial anisotropies depending on what
kinds of orbital is considered and the symmetry is in general 
not fully isotropic in the orbital space. These differ from the typical
case of spin systems, where the spin anisotropy is zero or not so strong
unless the spin-orbit interaction is very strong. 
 Each orbital order
possesses a unique property and exploring such uniqueness is an important
issue of the condensed matter theories and experiments.

Recently, Pr-based compounds Pr$T_2X_{20}$ ($T$=Ir,Rh,Ti,V, and
$X$=Zn,Al), so-called 1-2-20 compounds, have attracted great attention.
\cite{Onimaru0,Onimaru,Onimaru2,Sakai1,Sakai2,Matsubayashi,Shimura,Ishii1,Ishii2,Matsushita}
In these compounds, each Pr ion has (4f)$^2$ electron configuration. Its
ground state under the crystalline electric field (CEF) is a non-Kramers
doublet,\cite{Iwasa} and this doublet couples with conduction electrons.
This is a typical situation that the two-channel Kondo effects take
place. Thus, it is expected that these compounds show
exotic properties due to the two-channel Kondo effects.\cite{Sakai1} Nevertheless, the
doublet is not completely screened and all the
compounds exhibit an orbital order below $\sim 1$ K. 
Surprisingly, some of them show even superconductivity at very 
low temperature in the ordered phase.\cite{Onimaru0,Onimaru2,Sakai2,Matsubayashi} 

Since the first-excited CEF multiplet is located at $\sim 30$ K\cite{Onimaru0,Onimaru,Onimaru2,Iwasa,Sato} and the
phase transition occurs at $\sim 1$ K, it is likely that 
the phase transition in the 1-2-20 compounds is about the degrees of freedom
in the ground-state multiplet, {\it i.e.,} an ordering of the non-Kramers 
doublets. 
 The ground-state non-Kramers doublet is denoted as $\Gamma_3$ according
 to the standard notation of irreducible representations (irreps) of $T_d$
 group.\cite{irreps} 
Within the $\Gamma_3$ doublet states, two quadrupole operators and one octupole
operator have finite matrix elements. The former belong to $\Gamma_3$
irrep, while the latter $\Gamma_2$.
Thus, the (local) 
order parameter of the orbital order is either $\Gamma_3$
quadrupole or otherwise $\Gamma_2$ octupole. There has been no direct
evidence for which type of order is realized, { but results of 
the neutron scattering\cite{Sato} and the ultrasound experiments\cite{Ishii1,Ishii2}
 suggest a quadrupole order and no evidence for octupole or magnetic
 dipole order has been observed so far.}

Quadrupole orders were intensively discussed for f-electron system 
 CeB$_6$.\cite{Kura,CB6-shiina,CB6-thalmeier} { and another example
 TmTe\cite{Matsumura,ShiinaTmTe}. The CEF ground state of Ce (Tm)
 ion is the $\Gamma_8$ quartet in $O_h$ symmetry, where it has a
 4f$^{1}$(4f$^{13}$) 
 electron configuration in contrast to the case of the Pr 1-2-20
compounds. Despite this difference, they also exhibit a quadrupole
 order. On the basis of a localized model, the results by mean-field
 approximations are qualitatively consistent with the experimental ones.\cite{CB6-shiina,ShiinaTmTe}
} 
Several other systems also have a non-Kramers $\Gamma_3$ doublet ground
 state, and PrPb$_3$\cite{PrPb3}, PrInAg$_2$\cite{PrInAg}, and
 PrMg$_3$\cite{PrMg3} are examples. In the previous studies, experimental group tried fitting
 their experimental data by a mean-field approximation of the CEF model,
 for example, in PrRh$_2$Zn$_{20}$\cite{Onimaru2},
 PrPb$_3$,\cite{Tayama} and PrIr$_2$Zn$_{20}$.\cite{OnimaruP2}
 In this paper, we extend their analysis
and provide detailed theoretical account of $\Gamma_3$
quadrupole ordering. We have also discovered unusual
properties related to this order and unveiled their origin.

 The main purpose of this
 paper is to provide a basic understanding of the $\Gamma_3$ quadrupole
 order and the nature of its transition, rather than to make quantitative
 comparisons  with experimental data. 
Various points need careful analysis.
These include (i) 
 constraints on quadrupole ordered states by general arguments, (ii)
 excitation spectra in ordered states, and (iii) unusual criticality
 even within mean-field approximation. To this end, we employ the
 simplest model for inter-site interaction relevant to the quadrupole
 order and ignore all other
 interactions that might be present in the real systems.

In this paper, we will carry out a detailed theoretical analysis on
the $\Gamma_3$ quadrupole order in a diamond-lattice model with a minimal
quadrupole-quadrupole interaction on the basis of the CEF level scheme
relevant to 
 PrIr$_2$Zn$_{20}$ as an example. The model shows
quadrupole anisotropy and this is general for the $\Gamma_3$ degrees of freedom in
cubic symmetry. 
We will discuss it in detail, 
and determine and examine the mean-field
phase diagram in Sect. \ref{sec2}. We will then in Sect. \ref{Excitations}
study excitation spectra based on
spin-wave type 
approximations and clarify unusual critical behaviors found in this system. In Sect. \ref{Discussions}, we
will discuss some experimental results of the 1-2-20 compounds. We will 
also study  possible
 couplings of quadrupoles with other degrees of freedom including
other multipole moments and phonons. Section \ref{Summary} is a summary of the present paper.


\section{A Basic Model and Mean-field Approximation}\label{sec2}
In this section, we will first discuss the effect of single-ion
anisotropy and its effect on 
quadrupole orders. Then, we will analytically analyze 
the mean-field ground state for zero magnetic field. In the final part
of this section, we will show numerical results and discuss
the temperature vs. magnetic-field phase diagram.

\subsection{Crystalline-electric-field Hamiltonian}\label{CEF}
We start with investigating a CEF model. Each Pr ion has 
total angular moment $J=4$ and feels the CEF with the local $T_d$ point group symmetry. The
CEF states are represented by the irreps of 
 the $T_d$ group. The nine 
 states in the $J=4$ multiplet are splitted by the CEF as shown by 
\begin{eqnarray}
H_{\rm CEF}\!\!\!\!\!\!&=&\!\!\!\!\!\!\sum_i\Big\{\epsilon_2 |{\bf
 Q}(i)|^2-\epsilon_3\Big[Q_z^3(i)-3\overline{Q_z(i)Q_x^2(i)}\Big]
 \Big\}.\ \ \ \ \label{eqHcefQ}
\end{eqnarray}
 Here, $\epsilon_{2,3}$ are constants and ${\bf Q}(i)$ is the quadrupolar operator at the site $i$ which is represented by the Stevens
operators\cite{Stev} as ${\bf Q}(i)=(Q_z(i),  
Q_x(i))=(O^0_2(i),\sqrt{3}O^2_2(i))/8$. In terms of dipole
operators $J_{x,y,z}(i)$ for the $J=4$ manifold,
$O_2^0(i)=2J_z^2(i)-J_x^2(i)-J_y^2(i)$ and $O_2^2(i)=J_x^2(i)-J_y^2(i)$.
The bar denotes a cyclic permutation of three operators:
$ 
3\overline{AB^2}\equiv
AB^2+BAB+B^2A$.
With this normalization, the operator ${\bf Q}(i)$ within the $\Gamma_3$
subspace is represented by the two Pauli matrices ${\bf Q}\sim
(\tau_z,-\tau_x)$ with the basis states listed in Appendix\ref{Wavefunc}. 
The relation between Eq. (\ref{eqHcefQ}) and the conventional form of
the CEF Hamiltonian (\ref{eqHcef0}) is summarized in Appendix
\ref{RelationHcef}. The advantage in using Eq. (\ref{eqHcefQ}) is that
one can easily understand the single-ion anisotropy in the quadrupole sector.

\subsection{Single-site anisotropy and quadrupole ordering}\label{singlesite}
The CEF Hamiltonian (\ref{eqHcefQ}) readily indicates that the anisotropic $\epsilon_3$ term
affects the quadrupole order parameters $Q_z$ and $Q_x$ {\it
differently}. This is related to the existence of $Z_3$ symmetry in
${\bf Q}$ space. 
We can
rewrite the $\epsilon_3$ term in Eq. (\ref{eqHcefQ}) as
\begin{eqnarray}
Q_z^3-3\overline{Q_zQ_x^2}&=& 
\frac{4}{3}\sum_{l=0}^2Q_l^3,
\end{eqnarray}
where we have omitted the site index $i$ and $Q_l={\bf n}_l\cdot {\bf Q}$ with
${\bf n}_l=(\cos 2l\pi/3,\sin 2l\pi/3)$. 
Any symmetry operations in cubic groups are
reduced to permutations of $(lmn)$ in the $Q_l$
space with $0\le l,m,n\le 2$. 
There exists an essential difference between $Q_z$ and $Q_x$. 
The inversion $Q_x\to -Q_x$ corresponds to the exchange of $Q_1$ and
$Q_2$ and this is a symmetry operation of $H_{\rm CEF}$. However, the
inversion $Q_z\to -Q_z$ is not a symmetry operation.
This is evident from the fact
that the $\epsilon_3$ term contains odd numbers of $Q_z$.

\begin{figure}[t]

\begin{center}
    \includegraphics[width=0.45\textwidth]{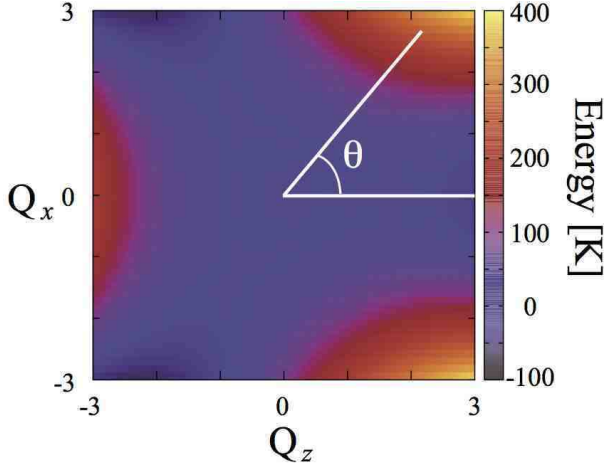}
\end{center}
\vspace{-4mm}
\caption{(Color online) Classical potential energy (single-ion
 anisotropy) due to Eq. (\ref{eqHcefQ}) in the quadrupolar space with 
$\epsilon_2=9.709$ K and $\epsilon_3=4.088$ K.
}
\label{fig-pot}
\vspace{-3mm}
\end{figure}

To understand this point more intuitively, we map the $\epsilon_3$ term
to a classical one.
This leads
\begin{eqnarray}
Q_z^3-3\overline{Q_zQ_x^2}
\to Q^3 \cos 3\theta,\label{classicalmap}
\end{eqnarray}
where we have introduced the polar coordinates ${\bf Q}$$=$$Q[\cos\theta,\sin\theta]$.\cite{Th}
Thus, it is apparent that $\theta$ and $-\theta$ are equivalent in energy, while
$\theta$ and $\pi-\theta$ is inequivalent as shown in Fig. \ref{fig-pot}.\cite{Potts}

This anisotropy in ${\bf Q}$-space imposes an important constraint 
on the symmetry of quadrupole ordering.
We will study this ordering in detail in Sect. \ref{Model}, but now
 briefly discuss the constraint due to the anisotropy.
The most important point is the impossibility of a pure 
antiferro $Q_x$ ($O_2^2$) order,  $\langle {\bf Q}(i)\rangle=(0,\pm Q)$
unless the Hamiltonian is finely tuned. 
This state is {\it unstable} because of the $\epsilon_3$
term. The single-ion anisotropy $H_{\rm CEF}$ favors the three
directions in the ${\bf Q}$ space: $\theta=2l\pi/3
(l=0,1,2)$
for $\epsilon_3>0$ and $\theta=(2l+1)\pi/3$ for $\epsilon_3 <0$. 
Thus, a uniform $\langle Q_z\rangle$ is induced, and its sign is 
determined by the sign of the anisotropy energy $\epsilon_3$. 
When $\epsilon_3>0$, $\langle Q_z\rangle <0$, while $\langle Q_z\rangle>0$
 when $\epsilon_3 <0$. In contrast to the antiferro $Q_x$ order, a pure 
antiferro or ferri $Q_z$ ($O_2^0$) order is possible; 
$\langle {\bf Q}(i)\rangle=(Q,0), (-Q',0)$.
Note, however, that, when $\epsilon_3<0$, the the ground state of $H_{\rm CEF}$ is not
$\Gamma_3$.
Although this argument is based on the classical mapping
(\ref{classicalmap}), the result is valid for the quantum  
Hamiltonian as will be discussed in Sect. \ref{AnalisisMF}, where we will
see similar anisotropic term appears in the free energy and in the
ground-state energy.

The difference between $Q_x$ and $Q_z$ orderings discussed above becomes
clearer if one notices the {\it real} symmetry in the ${\bf Q}$ space.
 As the energy landscape in Fig. \ref{fig-pot} shows, the ${\bf Q}$
 order parameter space has the trigonal symmetry $Z_3$, not the
 continuous O(2) symmetry. Any quadrupole ordering is actually the
 breaking of this $Z_3$ symmetry. 
The three-fold symmetry ($Z_3$) represents the equivalence of the three $(x,y,z)$ principle
axes of quadrupole moments in real space.

\begin{figure}[t!]

\begin{center}
    \includegraphics[width=0.5\textwidth]{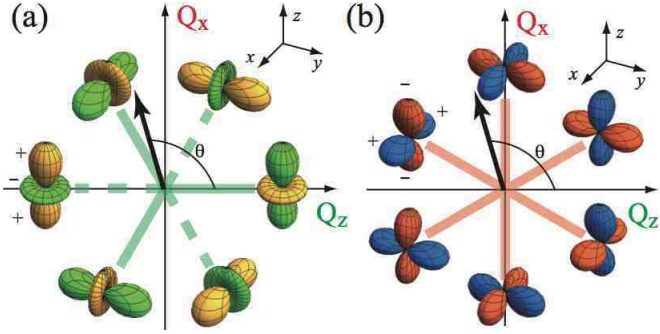}
\end{center}
\caption{(Color online) Quadrupole order parameters in $Q_z$-$Q_x$
 plane for (a) $O_2^0$ type and (b) $O_2^2$ type. $\pm$ indicates the
 sign of the charge distribution. {Note that the shapes of order
 parameters shown do not
 reflect the real charge density of f-electron orbitals, but just represent the
 symmetry of $\Gamma_3$ irreps.}

}
\label{fig-Q}

\end{figure}

To understand this, it is instructive to show the types of quadrupole order parameter in
the two-dimensional ${\bf Q}$ space. Figure \ref{fig-Q}
illustrates shapes of quadrupole moments in the ${\bf Q}$ space.
Namely $\theta=0$, $2\pi/3$,
and $4\pi/3$ correspond to quadrupole moments with $(3z^2-r^2)$, $(3x^2-r^2)$, and
$(3y^2-r^2)$, respectively, with $\theta+\pi$ corresponding to
those with opposite sign and $r^2=x^2+y^2+z^2$. Similarly, those for $(x^2-y^2)$,
$(y^2-z^2)$ and $(z^2-x^2)$ are at $\theta=\pi/2$,
$7\pi/6$, and $11\pi/6$, respectively with opposite sign for $\theta+\pi$. 


\subsection{A model for the Pr 1-2-20 compounds}\label{Model}
\begin{figure}[t!]

\begin{center}
    \includegraphics[width=0.45\textwidth]{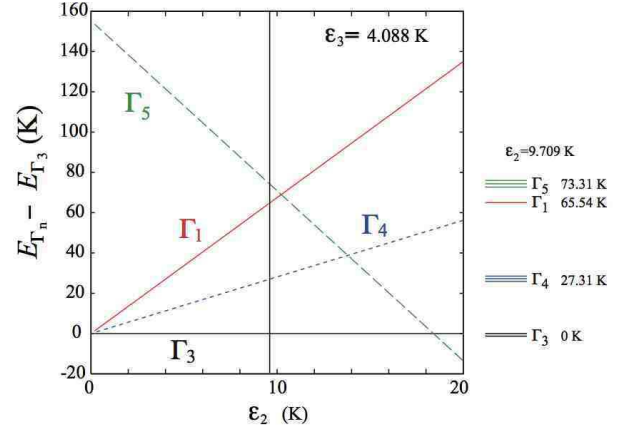}
\end{center}
\caption{(Color online) Single-ion energy level scheme determined by
 $H_{\rm CEF}$ for $\epsilon_3=4.088$ K. Vertical line represents
 $\epsilon_2=9.709$ K and the energy scheme there is shown in the right.}
\label{fig-cef}

\end{figure}

{
The model we discuss throughout this paper includes only 
degrees of freedom of localized f-electrons with 
quadrupole-quadrupole exchange interaction under zero or finite magnetic field. 
Although quite simplified, this is a natural model to qualitatively 
describe essential aspects of the Pr-based 1-2-20 compounds.
The transition temperature of the unidentified order in these systems is below
1 K, while the lowest CEF level above the $\Gamma_3$ ground state has
the excitation energy of $\sim 30$
K.\cite{Onimaru0,Onimaru,Onimaru2,Iwasa,Sato}  
This indicates that the doublet 
ground states play a dominant role in this transition. 
In the $\Gamma_3$ doublet, quadrupole and octupole moments are 
 the only active operators. 
Therefore, their orderings are the most natural candidate of the
transition. 
For example, a quadrupole order is suggested
 by anomaly observed in ultrasound experiments, and they also show 
 that the inter-site quadrupole-quadrupole interaction is antiferro for
 PrIr$_2$Zn$_{20}$\cite{Ishii1} and PrRh$_2$Zn$_{20}$.\cite{Ishii2} 
For PrTi$_2$Al$_{20}$, the quadrupole interaction is ferro\cite{Nakanishi} and this is
also consistent with the neutron scattering experiments.\cite{Sato}
These strongly suggest that the transition is an ordering of these
quadrupoles, and thus, we investigate this case in detail in this paper.

Concerning inter-site interactions, there also exist some magnetic exchange
 interactions, which excite the $\Gamma_3$ ground state to $\Gamma_4$
 and $\Gamma_5$  levels. 
They may modify, for example, the details of the temperature-magnetic
 field phase diagram, but the essential part of all the properties in 
this transition is described by our minimal model. In
this paper, we concentrate on
analyzing the simplest model with quadrupole exchange interactions 
 defined as }
\begin{eqnarray}
H\!\!\!\!&=&\!\!\!\!\!H_{\rm CEF}+\lambda\sum_{\langle i,j\rangle}
 {\bf Q}(i)\cdot {\bf Q}(j)
-\mu\sum_i{\bf H}\cdot {\bf J}(i).\ \ \ \ \ \label{H}
\end{eqnarray}
Here, we consider this Hamiltonian on the diamond lattice of Pr sites. 
 ${\bf J}(i)=(J_x(i),J_y(i),J_z(i))$ represents the magnetic dipole operator at the site
$i$. When the magnetic field ${\bf H}$ is applied, it couples with ${\bf J}$.
Here, the coupling constant is the ``magnetic moment'' (more precisely,
gyromagnetic constant) $\mu$, and 
$\mu=g\mu_B$ with $g=4/5$ for
$J=4$ multiplet and $\mu_B$ being the Bohr magneton. 
The quadrupole-quadrupole interaction $\lambda$ is set to $\lambda > 0$,
{\it i.e.}, antiferro, and the summation is over nearest-neighbor pairs. 
Although a ferro quadrupole order is suggested in
PrTi$_2$Al$_{20}$,\cite{Sakai1} we do not discuss the case for
$\lambda<0$ in detail in this paper. Some comments on ferro quadrupole
orders will be given in Sect. \ref{FQ}.

For $\epsilon_2$ and $\epsilon_3$ in $H_{\rm CEF}$, we choose the
representative values for the 1-2-20 compounds similar to those
determined by 
inelastic neutron scattering experiments for PrIr$_2$Zn$_{20}$:\cite{Iwasa}
$\epsilon_2=9.709$ K and $\epsilon_3=4.088$ K. 
This leads CEF level scheme shown in Fig. \ref{fig-cef}: $E_{3}=0$ K, $E_{4}=27.31$ K,
$E_{1}=65.54$ K, $E_{5}=73.31$ K.
The CEF level scheme 
 with fixed $\epsilon_3=4.088$ K is plotted as a function of
 $\epsilon_2$ in
Fig. \ref{fig-cef}, where we show the part where the ground state is
$\Gamma_3$ doublet.

{
It should be noted that conduction electrons in the 1-2-20 compounds is not
taken into account in Eq. (\ref{H}). We do not discuss the Kondo physics
or other properties originating from coupling with conduction electrons.\cite{Sakai1,Sakai2} 
One may regard our localized model
 as a renormalized effective one after tracing out the
conduction electrons. It is natural to expect that this simplification is
not 
serious if one considers symmetry breakings of local moments as shown in earlier works\cite{CB6-shiina,CB6-thalmeier,ShiinaTmTe}.
}

It would be helpful to present one of our main results before explaining 
details of analysis. It is the mean-field phase diagram in the parameter 
space of temperature $T$ and magnetic field ${\bf H}$. Figure
\ref{fig-phase} shows the results for ${\bf H}$ along three
high-symmetry directions.
In the followings, we will discuss the nature of each phase in details.

\subsection{Analysis of ground state for zero magnetic field} \label{AnalisisMF}
In this subsection, we investigate the ground state for ${\bf H}={\bf 0}$ by a
mean-field approximation. For ${\bf
H=0}$, relevant Hilbert space at one site is reduced to three states, $\Gamma_3$
doublet and
$\Gamma_1$ singlet. 
The other states are decoupled since there are no quadrupole matrix elements
between $\Gamma_{1,3}$ states and $\Gamma_{4,5}$ states, and thus, we can safely neglect them.

As shown in Appendix \ref{app-T0E}, the ground-state energy is given by
Eq. (\ref{E0}).
We want to search for small $z\lambda/E_1$ a solution that minimizes
Eq. (\ref{E0}). Here, $z$ is the number of the nearest-neighbor sites
($z=4$ for the diamond lattice). 
In the limit of $E_1\to \infty$, 
the moment modulus $|\langle {\bf Q}\rangle|$ approaches 1 on $A$ and
$B$ sites, and the anisotropy energy vanishes. Therefore, in this limit, 
any antiferro quadrupole order $\langle {\bf Q}^B\rangle=-\langle {\bf
Q}^A \rangle$ with $|\langle {\bf Q}^{A,B}\rangle|=1$ is the ground
state irrespective of the direction of $\langle {\bf Q}^{A,B}\rangle$. 
Here, we refer to this order as ``antiferro,'' but note that the
ordering wave vector is ${\bf q}={\bf 0}$. This is because the unit cell
of the diamond lattice contains both $A$ and $B$ sublattices.
This degeneracy is lifted in the order of $(z\lambda/E_1)^3$ as we will
show  below.
When $z\lambda/E_1$ is finite but small,  $|\langle {\bf Q}^{A(B)}\rangle|=1+\delta
q^{A(B)}$, $\theta_A=\theta + \delta \theta$ and $\theta_B=\theta + \pi - \delta \theta$.
$\delta \theta$ represents the deformation of the antiparallel alignment
of 
$\langle {\bf Q}^A\rangle$ and $\langle {\bf Q}^B\rangle$, 
 which will turn out to be important. 
To explicitly show the anisotropic term in the energy, we
minimize this energy with respect to $\delta q^A$, $\delta q^B$ and
$\delta \theta$ for a given value of $\theta$. 
In the leading order, the optimized values of the parameters are 
\begin{eqnarray}
(\delta q^A, \delta q^B)\!\!\!\!\!&\simeq&\!\!\!\!\! \frac{35}{2} \Big(\cos^2
 \frac{3\theta}{2}, \ \sin^2 \frac{3\theta}{2}\Big) \Big(\frac{z\lambda}{E_1}\Big),   \label{dqA} \\
\delta \theta \!\!\!\!\!&\simeq&\!\!\!\!\! -\frac{105}{16} \sin 3\theta \Big(\frac{z\lambda}{E_1}\Big).\label{deta1}
\end{eqnarray}
Higher order corrections in Eqs. (\ref{dqA}) and (\ref{deta1}) affect
the ground-state energy
 (\ref{E0}) in the order $(z\lambda/E_1)^5$, and thus can be safely neglected.
Using Eqs. (\ref{dqA}) and (\ref{deta1}), we obtain 
\begin{eqnarray}
\frac{E_{\rm mf}^{\rm gs}}{E_1}\!\!\!&=&\!\!\! - \Big(\frac{z\lambda}{E_1}\Big) -
 \frac{35}{4}\Big(\frac{z\lambda}{E_1}\Big)^2 \nonumber\\
&+&\!\!\! \frac{35}{256}\Big(29+35\cos
6\theta\Big)\Big(\frac{z\lambda}{E_1}\Big)^3 +\cdots. \label{E0_2}
\end{eqnarray}
From Eq. (\ref{E0_2}), one can see that $\partial E_{\rm mf}^{\rm gs}/ \partial \theta=0$ at $\theta=n\pi/6$ with
$n$ being integers and these values of $\theta$ correspond to the order
parameters depicted in Figs. \ref{fig-Q}(a) and (b). 
For sufficiently small $z\lambda/E_1$, odd $n$ solutions [Fig. \ref{fig-Q}(b)]
have a lower energy than even $n$ solutions, and thus the ground state is six-fold
degenerate with $|\langle {\bf Q}^A\rangle|=|\langle {\bf Q}^B\rangle|$.  
The primary order parameter is antiferro quadrupole of $Q_x(O_2^2)$
type, and as discussed in Sect. \ref{singlesite}, this is generally
accompanied by 
ferro quadrupole of $Q_z(O_2^0)$. This is similar to the case of
parasite ferromagnetism in antiferromagnets with Dzyaloshinskii-Moriya
interaction. 
The amplitude of the parasite $Q_z$ moment is determined by the 
deformation angle, $\delta \theta$ between ${\bf Q}^A$ and $
{\bf Q}^B$. It is small and of the order $z\lambda/E_1$, whereas the
amplitude of primary order parameter is $|Q_x|\sim 1$.

The six-fold degeneracy of the ground state is due to $Z_3\otimes Z_2$
symmetry of the lattice. $Z_3$ symmetry is about the equivalence of the 
three principle axes
$(x,y,z)$. $Z_2$ is the symmetry between the two sublattices. 
The emergence of the secondary order in the $Q_z$ component is
consistent with the analysis based on the single-site CEF anisotropy
discussed in Sect. \ref{singlesite}.

When the intersite coupling $\lambda$ increases, the quadrupole order in
the ground state changes from the $Q_x$ antiferro order to the $Q_z$
antiferro order through a first-order transition.  One can see this
behavior in the fact that the $\cos 6\theta$ term in the variational
energy (\ref{E0_2}) has a negative coefficient in the order
$(z\lambda/E_1)^4$, {\it i.e.}, the sign is opposite to that in the
leading order $(z\lambda/E_1)^3$. To examine the change in $\langle {\bf
Q}^{A,B}\rangle$
 quantitatively, we have numerically solved the mean-field equations and
 found that the transition occurs at $z\lambda/E_1\sim 0.0375$.


\subsection{Non-linear Zeeman term for quadrupole moments}
Let us  briefly discuss  effects of magnetic fields on $\Gamma_3$ degrees
of freedom. 
For small magnetic field, its coupling to the
quadrupole is calculated by the second-order perturbation in ${\bf H}$, in
which the intermediate virtual states are excited magnetic $\Gamma_4$
and $\Gamma_5$ states, and is given as  
\begin{eqnarray}
H_{Q}
 \!\!\!\!\!&=&\!\!\!\!\!-\alpha\Big[(2H_z^2-H_x^2-H_y^2)Q_z+\sqrt{3}(H_x^2-H_y^2)Q_x\Big],\
 \ \ \ \ 
\label{Hmag2}\\
\alpha \!\!\!\!\!&=&\mu^2\Big(\frac{7}{3E_{4}}-\frac{1}{E_{5}}\Big),
\end{eqnarray}
where $\alpha>0$ for our choice of the CEF level scheme.
This coupling leads to the effects of applied magnetic field on
quadrupole orders. 

Related to the cubic crystal symmetry or equivalently
$Z_3$ symmetry in the ${\bf Q}$-space for ${\bf H}={\bf 0}$, there exist three
equivalent ordered states, and they form a multi-domain structure.
These three states correspond to the angle $\theta=\pi/2$, $7\pi/6$, and
$11\pi/6$.
 Magnetic field ${\bf H}$ favors some of the three domains and disfavor
 the others, depending on the field direction, and thus controls the
 domain structure. When ${\bf H} \parallel$ [110], the domain of
 $\theta=\pi/2$ is favored.
This domain corresponds to $\sim (x^2-y^2)$ state.
In a similar way, for ${\bf H} \parallel$ [001] two domains
$\theta=7\pi/6: (y^2-z^2)$ and $11\pi/6: (z^2-x^2)$ are selected, since the magnetic field
favors positive $Q_z$. For [111] direction, the quadratic terms in 
(\ref{Hmag2}) vanishes and the leading effect is the coupling to octupole,
$
\sim H_xH_yH_z T_{xyz}
$.
Thus, $T_{xyz}$ octupole moment is induced for ${\bf H} \parallel$ [111] direction, where the
degeneracy of the three domains of $Q_x$-type antiferro quadrupole
states is not lifted.
\begin{figure}[t!]

\begin{center}
    \includegraphics[width=0.45\textwidth]{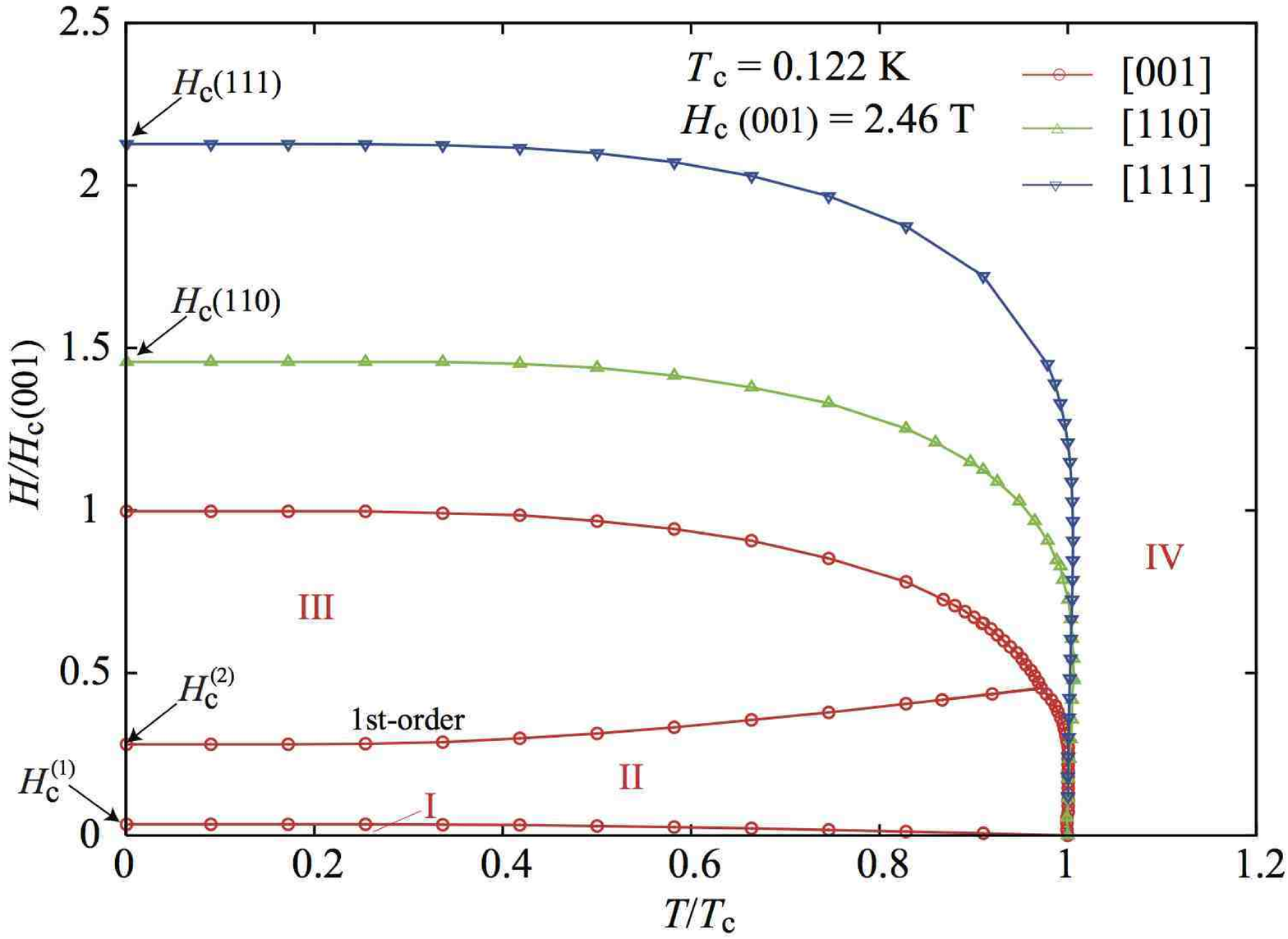}
\end{center}
\vspace{-3mm}
\caption{(Color online) Temperature-magnetic field phase diagram for
 ${\bf H} \parallel$
 [001], [110], and [111].}
\label{fig-phase}

\end{figure}
\begin{figure}[t!]

\begin{center}
    \includegraphics[width=0.45\textwidth]{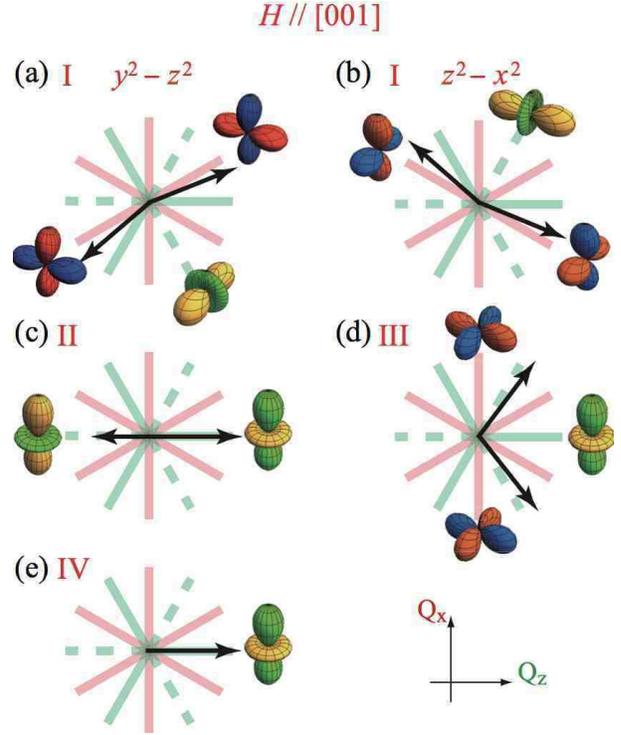}
\end{center}
\caption{(Color online) Schematic pictures of quadrupole orders
 for ${\bf H} \parallel$ [001]. (a) and (b) two domains for phase-I. 
(c) $Q_z$ antiferro quadrupole state in phase-II. (d) canted state
 for phase-III. (e) high-field polarized state.}
\label{fig-phaseconfig}

\end{figure}

Note that the
above discussions do not include the contribution of induced magnetic multipoles that
originates from the hybridization between $\Gamma_3$ and the excited
states. 
When the magnetic field is weak,  the induced magnetic parts 
 do not play important
roles in selecting domains and the direct effects in the $\Gamma_3$ sector dominates. 

\subsection{Temperature-magnetic field phase diagram} \label{THphase}
\begin{figure}[t!]

\begin{center}
    \includegraphics[width=0.45\textwidth]{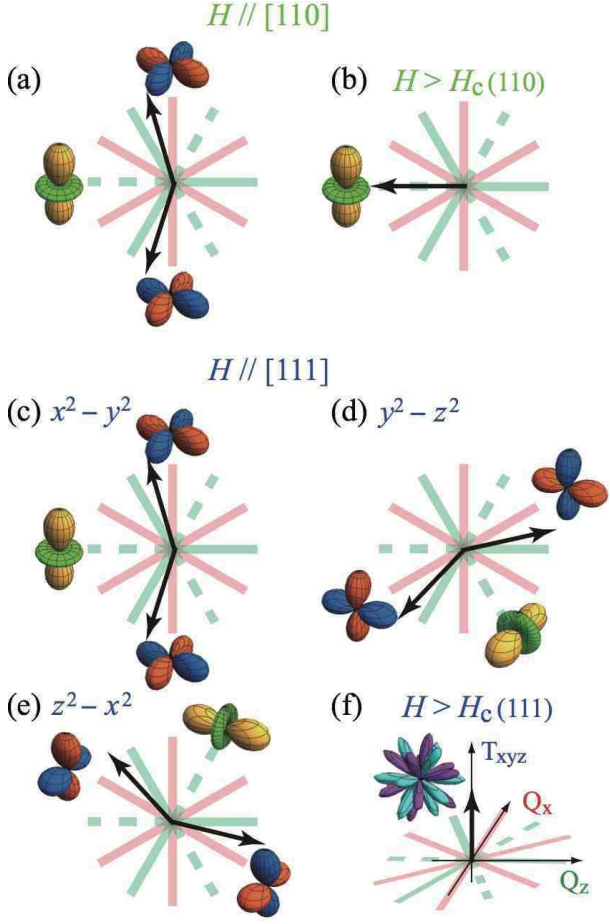}
\end{center}
\caption{(Color online) Schematic pictures of the quadrupole orders
 for ${\bf H} \parallel$ [110] and [111]. (a) antiffero
 quadrupole state for [110] direction. (b) high-field state for [110]
 direction. (c),(d),(e) three domains in antiffero quadrupole state for
 [111] direction. (f) high-field state for [111] direction. In (f), 
the third axis represents octupole $T_{xyz}$.}
\label{fig-phaseconfig2}

\end{figure}
In this subsection, we will discuss the $T$-$H$ phase diagram shown in
 Fig. \ref{fig-phase} in details
 and
 generalize
 the analysis in the previous subsection. To this end, we study
 the full model (\ref{H}).
The phase diagram
shows several ordered phases under magnetic field and we will examine
the nature of these phases.

Figure \ref{fig-phase} shows $T$-$H$ phase diagrams determined by the
mean-field analysis and ${\bf H}$ is 
parallel to each of the three high-symmetry directions. The intersite coupling constant $\lambda$ is
set to $\lambda=0.03$ K with $z=4$ neighbors. This leads to the transition
temperature $T_c\simeq 0.122$ K 
at ${\bf H}={\bf 0}$
and the critical magnetic field 
along ${\bf H} \parallel $[001]: $H_c(001)\simeq 2.46$ T, $H_c(110)\simeq
1.46H_c(001)$, and $H_c(111)\simeq 2.13H_c(001)$ at $T=0$, which 
roughly correspond to the experimental data.\cite{Ishii1} 
{Anisotropy in the critical field
strength is mainly determined by the non-linear Zeeman effect
(\ref{Hmag2}). The critical field is largest for ${\bf H} \parallel$ [001], then next 
for ${\bf H} \parallel$ [110], and smallest for ${\bf
H} \parallel$ [111].
For [111] direction, the effect (\ref{Hmag2}) vanishes and as discussed before
the critical field is determined by the octupole-field coupling as far 
as the magnetic field is not so strong.}

First, let us investigate
the case of ${\bf H} \parallel$ [001]. The phase diagram shows three ordered
phases I, II, and III as shown in Fig. \ref{fig-phase}. The phase-I is
 a basically antiferro quadrupole state with $Q_x$ type order with small
$Q_z$ ferro quadrupole components, and this is discussed in the
previous subsection. For finite $|{\bf H}|$, there are two
stable domains characterized by $(y^2-z^2)$
and $(z^2-x^2)$ type symmetries as illustrated in
Figs. \ref{fig-phaseconfig}(a) and (b), where two arrows represent order
parameters for two sublattices. With increasing the
field, 
the two arrows rotate clockwise in (a), while counterclockwise in 
(b). { This is accompanied by variations in the amplitude of
the quadrupole moments $|{\bf Q}^A|$ and
$|{\bf Q}^B|$. This is a natural way to gain the Zeeman energy
(\ref{Hmag2}), keeping the (almost) antiparallel configuration.}
At $|{\bf H}|=H_c^{(1)}$, there is a second-order transition to the
phase-II, where the two arrows point antipalallel along $Q_z$ direction
with different magnitudes as shown in Fig. \ref{fig-phaseconfig} (c). 
In the phase-II, the two domains in the phase-I are merged to the same
domain, and this is the unique stable domain. 
Further increasing magnetic field leads
 to a first-order transition to the phase-III at $|{\bf H}|=H_c^{(2)}$, 
where ${\bf Q}$ shows a canted order. 
The $Q_x$ component exhibit an antiferro order, while $Q_z>0$ ferro
order, as shown in Fig. \ref{fig-phaseconfig} (d).
With increasing magnetic field, the antiferro component decreases, while
the ferro component increases. 
Finally, there appears a second-order transition at $|{\bf H}|=H_c(001)$
from the canted phase-III to the polarized phase IV . In the phase-IV, the order parameter
points to $Q_z$ direction on both of $A$ and $B$ sites. The phase-IV is
smoothly connected to paramagnetic state above $T_c$.

Secondly, let us consider the case of ${\bf H} \parallel$ [110]. In this
case, there is only one ordered phase, and the stable domain is also
unique. This phase has an antiferro order of $(x^2-y^2)$-type quadrupole
component accompanied with ferro $Q_z<0$. This is illustrated in
Fig. \ref{fig-phaseconfig2} (a), and this coincides to one of the stable
domains in the phase-I when ${\bf H} \parallel$ [100] or [010].
With increasing field,
 the $Q_z$ components increase and finally the quadrupole
moments align along $-Q_z$ direction at $|{\bf H}|=H_c(110)$ through a second-order transition. 

Thirdly, for ${\bf H} \parallel$ [111], there is also only one ordered
phase but three domains are all stable. One domain exhibits negative $Q_z$ as
illustrated in Fig. \ref{fig-phaseconfig2} (c), while
the other two  have positive net $Q_z$ as illustrated in
Figs. \ref{fig-phaseconfig2} (d) and (e). Increasing $|{\bf H}|$
suppresses $Q_z$ for all the three, and, as discussed in the previous
subsection, induces $T_{xyz}$ octupole moment.
The transition to the high-field phase at $|{\bf H}|=H_c(111)$ is also second order. In the
high-field phase, no
quadrupole moment exists, while there remains an induced $T_{xyz}$
octupole moments as shown in Fig. \ref{fig-phaseconfig2} (f).

\subsection{Physical quantities}
\begin{figure}[t!]

 \begin{center}
    \includegraphics[width=0.4\textwidth]{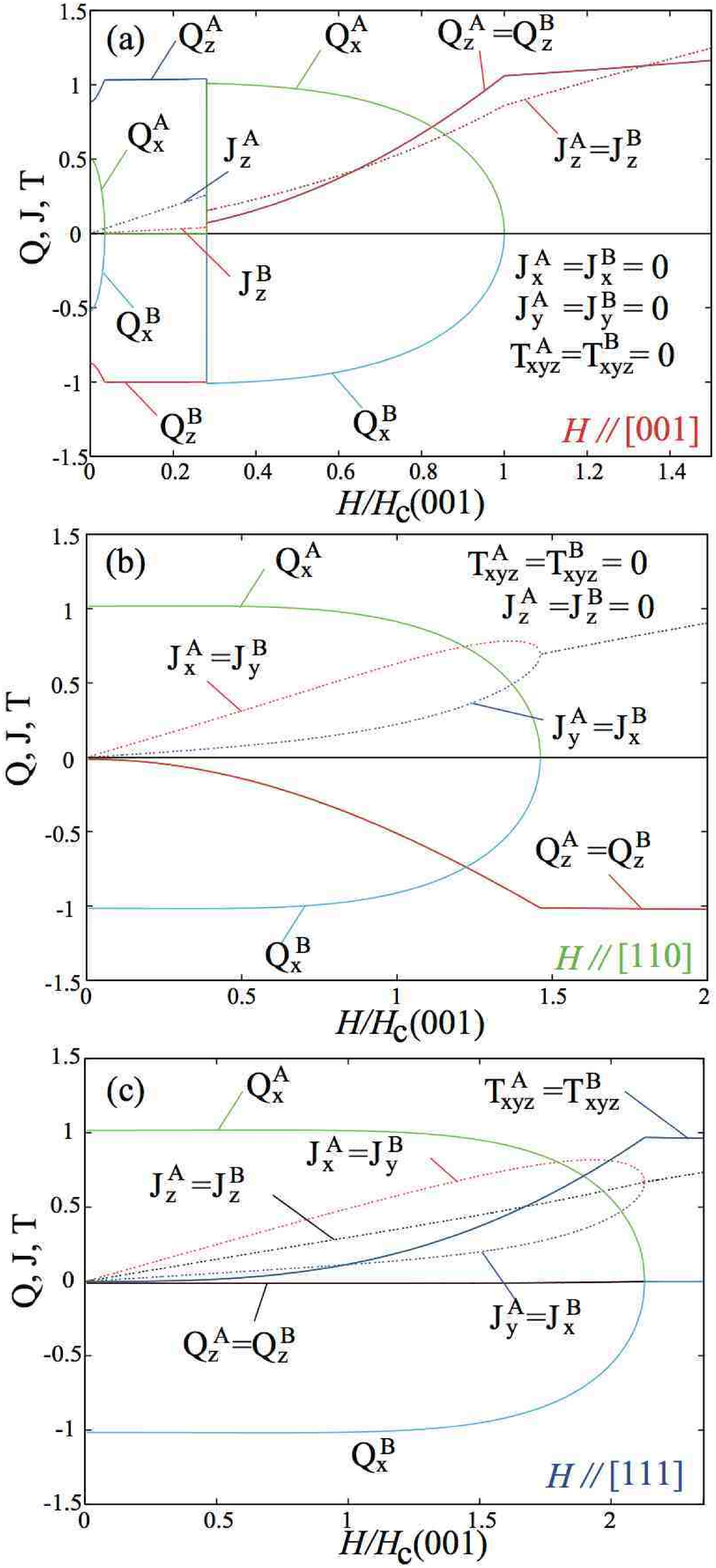}
 \end{center}
\caption{(Color online) Order parameters vs. magnetic field for (a)
 ${\bf H} \parallel$ [001], (b)  ${\bf H} \parallel$ [110], and (c)
  ${\bf H} \parallel$ [111].
}
\label{fig-op}

\end{figure}

In this subsection, we show the details of magnetic field dependence of
multipoles 
${\bf Q}$, ${\bf J}$, and $T_{xyz}$. Figure \ref{fig-op} shows these quantities at $T=0$ and for three directions of the
magnetic field.  The results for ${\bf H} \parallel$ [001] 
are shown  
in Fig. \ref{fig-op} (a), where the domain of $(y^2-z^2)$-type is chosen
for the phase-I. For ${\bf H} \parallel$ [110], the stable domain is
unique, {\it i.e.}, $(x^2-y^2)$-type, and Fig. {\ref{fig-op}}(b) shows multipole moments along ${\bf
H} \parallel$ [110]. The results for ${\bf H} \parallel$ [111] are shown 
in Fig. \ref{fig-op}(c), and the $(x^2-y^2)$-type domain is chosen for
the ordered phase. 
Since we have already discussed the variations of ${\bf Q}$ in Sect. \ref{THphase}, we here 
discuss a physical origin of induced magnetic moments ${\bf J}$. 

To discuss what kinds of multipoles are induced in the presence of both
magnetic field and the order parameter ${\bf Q}$, a
group theoretical argument is very useful.\cite{CB6-shiina} In Appendix \ref{direct}, a
list is shown for the reduction of several products of two irreps. 
This helps understanding the coupling of ${\bf J}$ to ${\bf H}$ and
${\bf Q}$, and it is sufficient to
notice that ${\bf Q}$ transforms as $\Gamma_3$ representation, while
${\bf J}$ and ${\bf H}$ as $\Gamma_4$. 
The lowest-order local coupling
that includes the three quantities is 
\begin{eqnarray}
\!\!\frac{Q_z-\sqrt{3}Q_x}{2}J_xH_x+
\frac{Q_z+\sqrt{3}Q_x}{2}J_yH_y-Q_zJ_zH_z.\ \label{QHJ}
\end{eqnarray}
This indicates that antiferromagnetic $J_z$ component is induced by the
antiferro $Q_z$ order for ${\bf H} \parallel$ [001] as shown in
Fig. \ref{fig-op}(a), 
while antiferromagnetic $J_x-J_y$ component is induced for ${\bf
H} \parallel$ [110] [Fig. \ref{fig-op}(b)] and [111] [Fig. \ref{fig-op}(c)] by the
antiferro $Q_x$ order. Note also that Eq. (\ref{QHJ}) holds for any $\Gamma_4$
operators by replacing ${\bf J}$ by them. 

Similarly, one can construct a
coupling between a magnetic $\Gamma_5$ operators, {\it e.g}., ${\bf T}^{\beta}$ octupole moment, and ${\bf H}$
and ${\bf Q}$, 
\begin{eqnarray}
\!\!\frac{\sqrt{3}Q_z+Q_x}{2}{T}^{\beta}_xH_x+
\frac{-\sqrt{3}Q_z+Q_x}{2}T^{\beta}_yH_y-Q_xT^{\beta}_zH_z.\ \label{QHJ2}
\end{eqnarray}
Although we do not show the results for $\Gamma_5$ operators,
 ${\bf T}^{\beta}$ octupole moments are induced as
discussed by Shiina {\it et al}.\cite{CB6-shiina}

Now, let us discuss changes in  the ground-state wavefunction
$|0\rangle_{A,B}$ 
with ${\bf H}$. 
With varying the strength of magnetic field, we numerically obtained the 
mean-field ground state on the $A$- and $B$-sublattices,
$|0\rangle_{A,B}=\sum_i\sum_n
a_{A,B}(\Gamma_i,n)|\Gamma_in\rangle_{A,B}$. 
$w_{A,B}(\Gamma_in)=|a_{A,B}(\Gamma_in)|^2$ is the occupation of the
basis state $n$ in the $\Gamma_i$ multiplet for the paramagnetic state
at ${\bf H}={\bf 0}$ in the ground state, and of course
$\sum_{in}w_{A,B}(\Gamma_in)=1$. 

\begin{figure}[t!]

\begin{center}
    \includegraphics[width=0.45\textwidth]{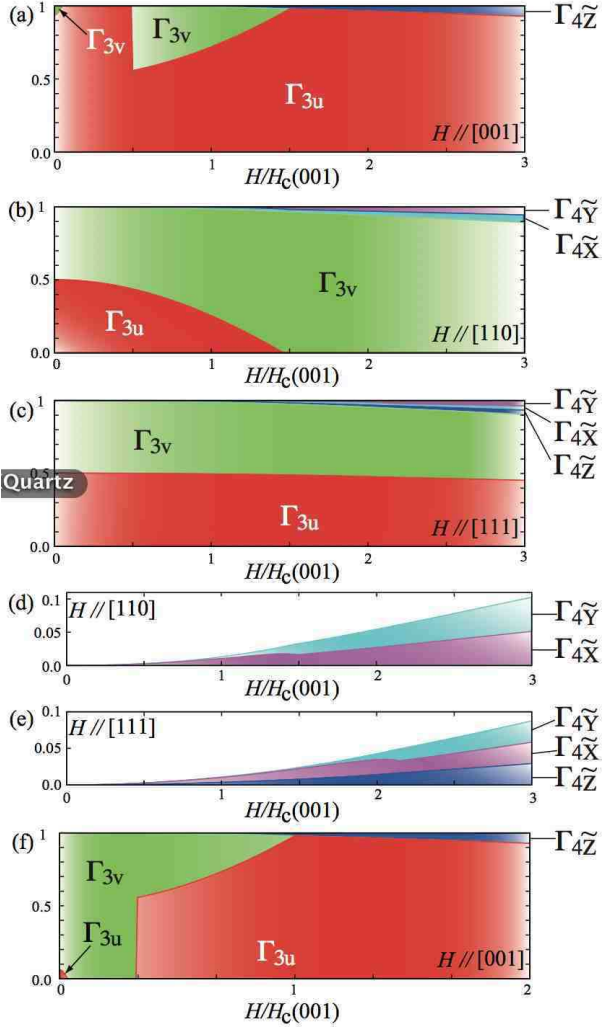}
\end{center}
\caption{(Color online) Weights of basis states in the $A$-site ground state for
 (a) ${\bf H} \parallel$ [001],  (b) ${\bf H} \parallel$ [110],  (c)
 ${\bf H} \parallel$ [111] with the $(x^2-y^2)$ domain.
Zoom up of $\Gamma_4$ weights for (d) ${\bf H} \parallel$ [110], (e) 
${\bf H} \parallel$ [111]. (f) Weights for $B$-sublattice for ${\bf H} \parallel$ [001].}
\label{fig-wf}

\end{figure}

Figures \ref{fig-wf}(a)-(e) show the occupation $w_A(\Gamma_i n)$ of the
five multiplets as a function of magnetic field strength for three field
directions. { The states not shown in each panel have occupation negligibly small or
exactly zero.} 
The occupation at the
$B$-sublattice is the same, $w_B(\Gamma_i n)=w_A(\Gamma_i n)$ except for
the case of $ {\bf H} \parallel$ [001],  for which $w_B(\Gamma_i n)$ 
 is shown in Fig. \ref{fig-wf}(f).

For ${\bf H} \parallel$ [001], two stable domains in the phase-I have
the same occupations $w$'s.
 In the phase-II the
wavefunction at the $A$-sublattice is almost pure $\Gamma_{3u}$ as in the high-field phase, since this is a collinear
order. At the $B$-sublattice, $\langle Q_z\rangle\sim -1$ and $\langle
Q_x\rangle=0$, and therefore, the wavefunction is almost a pure
$\Gamma_{3v}$ state.
In the phase-III, both $\Gamma_{3u}$ and $\Gamma_{3v}$ states have
large occupation at each sublattice and their hybridization yields a
canted configuration of ${\bf Q}$. With increasing field strength, one
 of the two states dominates, and correspondingly the ferro $Q_z$
 component increases.

\begin{figure}[h!]

\begin{center}
    \includegraphics[width=0.45\textwidth]{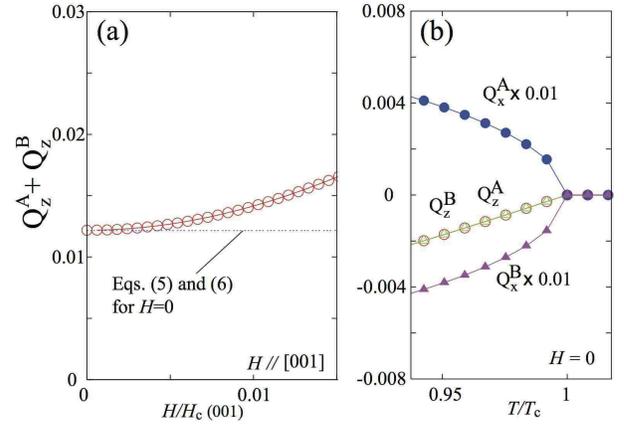}
\end{center}
\caption{(Color online) (a) Ferro $Q_z$ moment near ${\bf H}={\bf 0}$ for ${\bf H}
 \parallel$ [001]. The dotted line shows the value calculated by the
 perturbative analysis. (b) Temperature dependence of quadrupole moments near
 $T_c$ for ${\bf H}={\bf 0}$.}
\label{fig-op2}

\end{figure}

Let us check the amplitude of the secondary component and confirm that $\langle
Q_z\rangle\ne 0$ at ${\bf H}={\bf 0}$.  Figure
\ref{fig-op2}(a) shows the magnetic-field dependence of ferro $Q_z$
component near ${\bf H}={\bf 0}$ for $\bf H \parallel$ [001]. As one can
clearly see, the ferro component is indeed finite even at ${\bf H}={\bf 0}$
and the limiting value at ${\bf H}={\bf 0}$ agrees with that obtained by 
perturbative expressions (\ref{dqA}) and (\ref{deta1}). 

Temperature dependence of $Q_z$ near $T_c$ exhibits an evidence that the
ferro $Q_z$ moment is a secondary order parameter, and induced by $Q_x$
antiferro moment. 
Figure
\ref{fig-op2}(b) shows the temperature dependence of ${\bf Q}$ near
$T_c$. The primary order parameter $Q_x$ shows a typical mean-field
criticality, $|Q_x|\propto \sqrt{T_c-T}$, but as for the ferro $Q_z$
component, the temperature dependence is linear, $Q_z\propto
-(T_c-T)\propto -Q_x^2$.
 This fact is consistent with our early
analysis based on the third-order anisotropic term in $H_{\rm
CEF}$. 
As shown in Appendix \ref{Landau}, the local Landau free energy has a
third-order term. Combining this with the second-order term, the $Q_z$
part of the Landau free energy is given as 
\begin{eqnarray}
F_{Q_z}\sim \frac{1}{2\chi} Q_z^2 +\frac{\kappa_3}{\chi^2} Q_x^2 Q_z,
\end{eqnarray}
where the expression of $\chi$ and $\kappa_3$ are shown in Appendix
\ref{Landau}.
It is important that $Q_x^2$ have a static value in the ordered phase
below $T_c$, and therefore, $h_{Q_z}=-\frac{\kappa_3}{\chi^2}Q_x^2$ behaves as a
uniform conjugate field of $Q_z$. The induced moment is then
\begin{eqnarray}
Q_z=\chi h_{Q_z}=-\frac{\kappa_3}{\chi}Q_x^2.
\end{eqnarray}
This is consistent with the results in Fig. \ref{fig-op2}(b).
 This also explains why the induced $Q_z$
moment is ferro $Q_z$ and not antiferro $Q_z$.

\section{Excitations and Responses} \label{Excitations}
In this section, we will investigate ``spin'' wave excitations for $T=0$ and
static quadrupole susceptibilities in detail. 
\subsection{``Spin'' wave approximation at zero temperature}\label{SWA}
In this subsection, we will briefly explain the method of ``spin'' wave
approximation for a general exchange type
Hamiltonian.\cite{Joshi,KusuKura}
This analysis is equivalent to the equation of motion
method with decoupling\cite{CB6-thalmeier}  at $T=0$, and useful for analyzing
excitation spectra. Note, however, that because of the
nature of the approximation, the method is valid at low temperatures.

\subsubsection{Formulation}
We represent the fluctuation beyond the mean-field approximation by
using a 
set of bosons $\{a_{i l}\}$ and $\{b_{i l}\}$ and their conjugates.
Here, $a_{jl}^{\dagger}|0\rangle_A (l=1,\cdots,8)$ represents the $l$th 
mean-field excited state at the site $j$ in the 
 $A$-sublattice, while $b_{jl}^{\dagger}|0\rangle_B$ for $B$-sublattice, 
 and $|0\rangle_{A,B}$ represents the mean-field ground state at each site.
 The fluctuation
term in our model is 
\begin{eqnarray}
H_2\!\!\!\!\!\!&\equiv&\!\!\!\!\!\lambda\sum_{\langle i,j\rangle}\!
 \sum_{\mu=z,x}\Big[ Q^A_{\mu}(i)-\langle
 Q^A_{\mu}\rangle \Big]\Big[ Q^B_{\mu}(j)-\langle Q^B_{\mu}\rangle
 \Big],\ \ \ 
\end{eqnarray}
and in more general cases it is represented as 
\begin{eqnarray}
H_2\!\!\!\!\!\!&=&\!\!\!\!\!\!\!\sum_{m,\langle i,j\rangle}\!\!\!\lambda_m \Big[ O^A_{m}(i)-\langle
 O^A_{m}\rangle \Big]\Big[ O^B_{m}(j)-\langle O^B_{m}\rangle \Big].\ \ \ 
\end{eqnarray}
In terms of the introduced boson operators, this reads as
\begin{eqnarray}
H_2\!\!\!\!\!\!&=&\!\!\!\!\!\!\!
\sum_{m,l,l'\langle i,j\rangle}\!\!\!\lambda_m 
\Big[\big( O^A_{m}\big)_{l}a^{\dagger}_{i l}+{\rm h.c.}\Big]
\Big[\big( O^B_{m}\big)_{l'}b^{\dagger}_{j l'}+{\rm
h.c.}\Big].\ \ \ \ \ \  \label{Hfluc}
\end{eqnarray}
Here, $O^{A(B)}_{m}$ is a general operator at $A(B)$-sublattice appearing in the Hamiltonian
labeled by $m$ and $\big(O^{A(B)}_m\big)_{l}$
is the matrix element of $O^{A(B)}_m$ between the ground state and the $l$th excited
state for the $A(B)$-sublattice.  
Note that the linear term in the bosons vanishes due to the mean-field condition.
Combining on-site excitation energy $E^{A(B)}_l$ for the $l$th
excited state at the $A(B)$-sublattice, we obtain as a Hamiltonian for the bosons,
\begin{eqnarray}
H_{b}\!\!\!\!\!\!&=&\!\!\!\!\!\!\sum_{il}\Big(E^A_{l}a^{\dagger}_{i l}a_{i l}+E^B_{l}b^{\dagger}_{i l}b_{i l}\Big)
\nonumber\\
\!\!\!\!\!\!&+&\!\!\!\!\!\!
\sum_{m,l,l',\langle i,j\rangle}\!\!\!\!\lambda_m
\Big[\big( O^A_{m}\big)_{l}a^{\dagger}_{i l}+{\rm h.c.}\Big]
\Big[\big( O^B_{m}\big)_{l'}b^{\dagger}_{j l'}+{\rm h.c.}\Big]. \nonumber\\
&&\!\!\!\!\!\!\!\!\label{H2b}
\end{eqnarray}
This Hamiltonian is bilinear in the boson operators and therefore we can
diagonalize this  
by the Bogoliubov transformation 
to obtain its eigenenergies $\omega_{{\bf q}n}
(n=1,\cdots,16$ in our case) with
${\bf q}$ being the wavevector.

\subsubsection{Excitation spectrum at zero temperature}\label{SWA2}
Using the boson Hamiltonian introduced above, we now
calculate the evolution of excitation energy spectrum
with the variation of magnetic-field strength.
Diagonalization of $H_b$ is similar to random-phase approximation.
However, in contrast to the standard spin-wave theory for
magnets, all the local degrees of freedom are taken into
account.  The inter-site correlations are taken into account only through
the channels included in the microscopic Hamiltonian,
and they are ${\bf Q}$ operators in our case.


\begin{figure}[t!]

\begin{center}
    \includegraphics[width=0.5\textwidth]{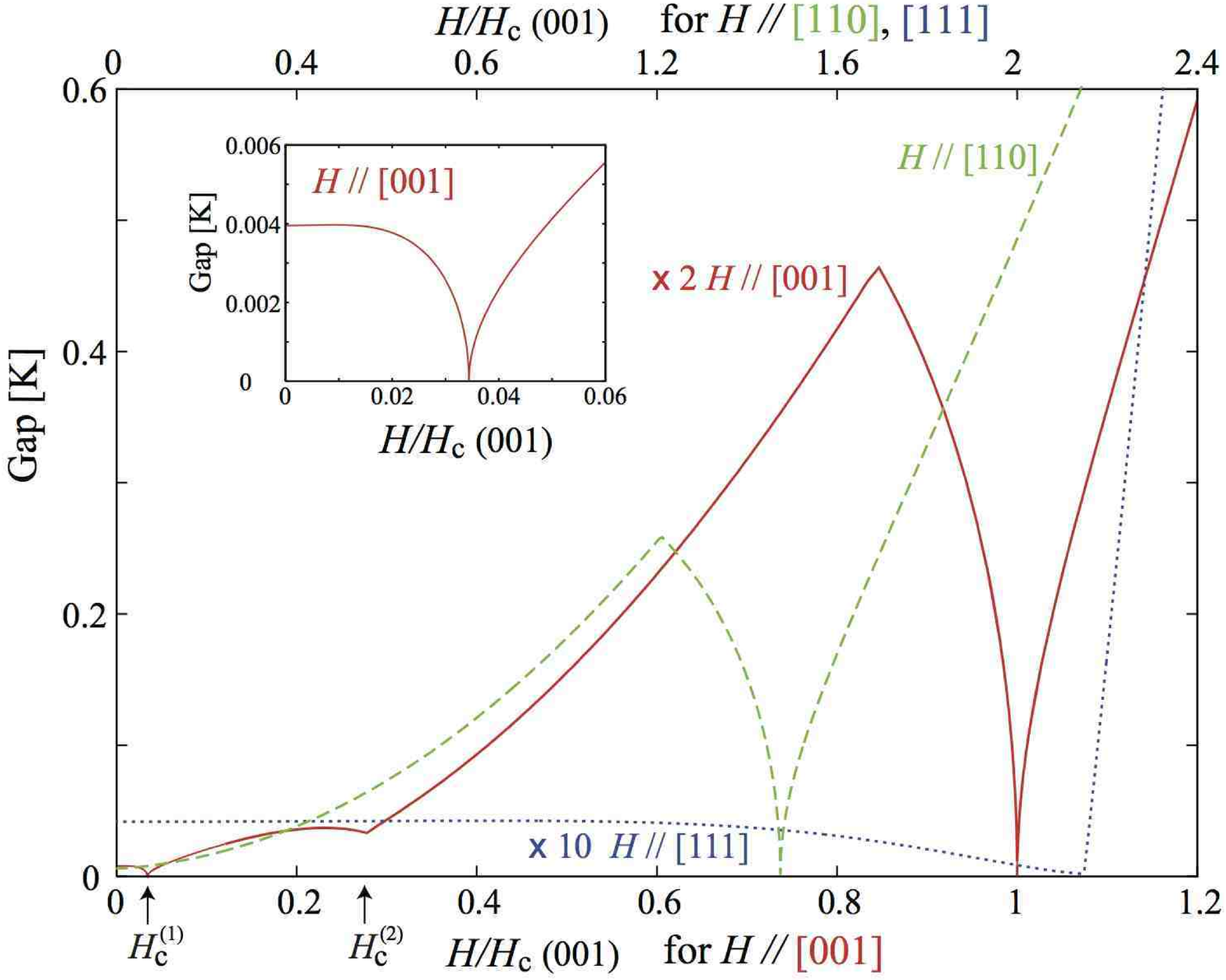}

\end{center}
\caption{(Color online) Spin wave excitation gap at ${\bf q}={\bf 0}$ for ${\bf
 H} \parallel$ [001] (solid line), for ${\bf H} \parallel [110]$ (dashed
 line), and for ${\bf H} \parallel [111]$ (dotted line). Beware that the
 scale of both axes is different depending on field direction. Inset: Zoom up
 of the low field regime for ${\bf H} \parallel [001]$.}
\label{fig-sw}

\end{figure}
Figure \ref{fig-sw} shows the lowest excitation gap as a function of
magnetic field for the three directions, where $H_c(001)$ is the critical
field between the phase-III and the phase-IV for ${\bf H} \parallel$ [001]. 
The excitations have an 
energy minimum at ${\bf q}={\bf 0}$ for all the parameter regimes. For each ${\bf H}$ direction,
the excitation becomes  gapless at the critical points 
 between the
ordered state and the high-field polarized state. This is very similar to
the case in transverse-field Ising systems. For ${\bf H} \parallel$
[001], 
there is an additional gapless point at $|{\bf H}|=H_c^{(1)}$ 
between the phase-I and the
phase-II. At the first-order transition point $|{\bf H}|=H_c^{(2)}$  
between the phase-II and the
phase-III, the magnetic field dependence of the gap exhibits a 
small jump (not visible in the scale in Fig. \ref{fig-sw}). 
One can see kinks near $H/H_c(001)\sim 0.8$ for ${\bf H} \parallel$ 
[001] and $H/H_c(001) \sim 1.2$ for ${\bf H} \parallel$ [110]. These are
not due to any phase transitions but due to level crossing between
different excited states.

Let us note that the scaling of 
the gap near the critical fields. For ${\bf H} \parallel$ [001] and
[110], the gap varies as $\propto |\delta H|^{1/2}$, where $\delta H$ is
the deviation from the critical field ${H}_c$ for each magnetic-field direction: $\delta {H}\equiv
{H}-{H}_c$. Interestingly, the gap for ${\bf
H} \parallel$ [111] varies as $|\delta H|^{3/2}$ for $H<H_c(111)$ and
 as $|\delta H|$ for $H>H_c(111)$. This leads to unusual divergence
in susceptibilities near the critical point for ${\bf H} \parallel$ [111].
We will  discuss this in more detail in Sects. \ref{chiQ} and \ref{GLsus}.

\begin{figure}[t!]

\begin{center}
    \includegraphics[width=0.5\textwidth]{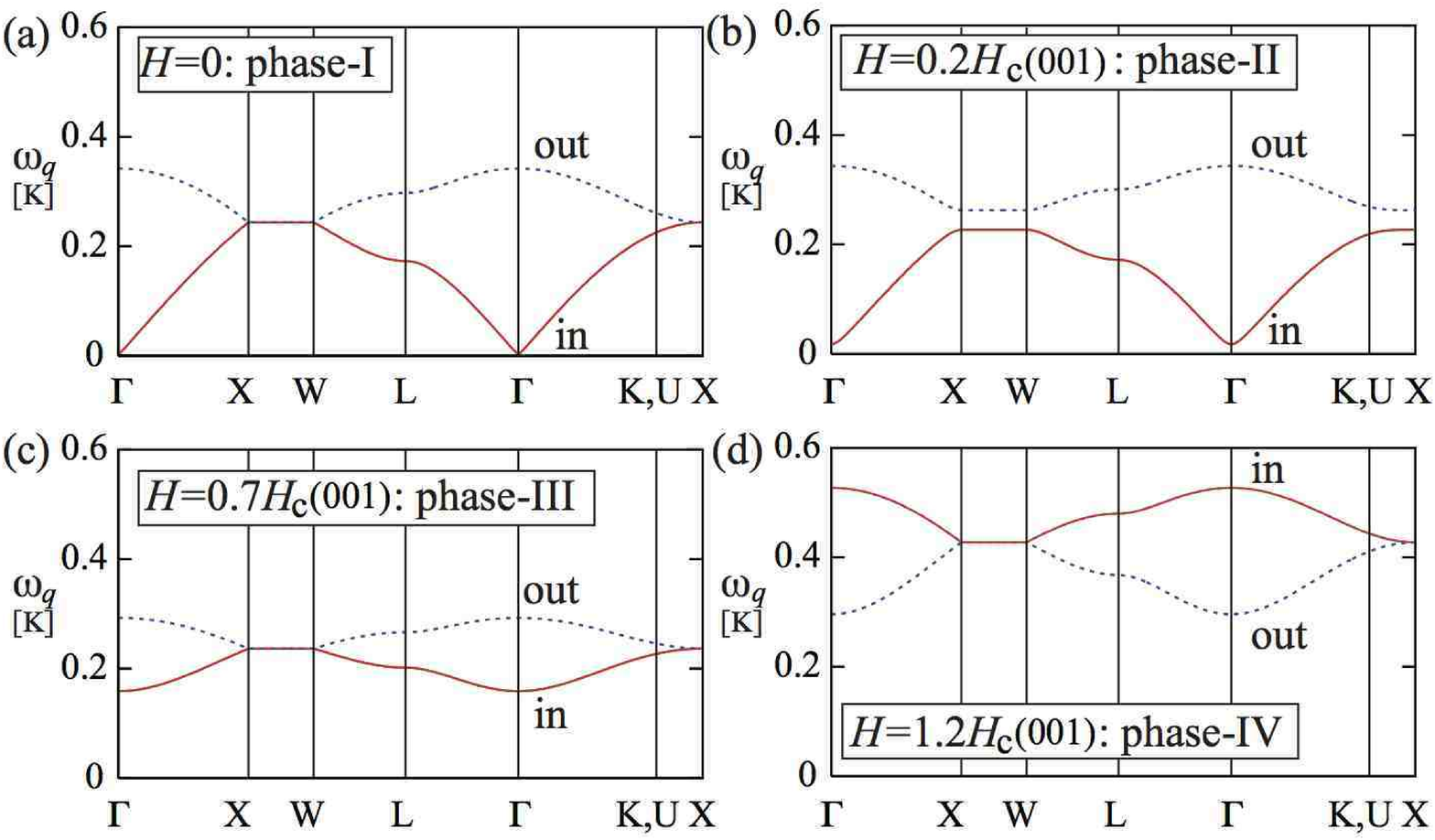}

\end{center}
\caption{(Color online) Excitation spectra for ${\bf H} \parallel$
 [001]. Full-red (dotted-blue) lines represent ``in'' (``out'')
 mode. See the text for the definition of the modes. (a) phase-I for
 ${\bf H}={\bf 0}$, (b) phase-II for $|{\bf
 H}|=0.2H_c(001)$, (c) phase-III for $|{\bf H}|=0.7H_c(001)$, and (d)
 phase-IV for $|{\bf H}|=1.2H_c(001)$.
Notations: $\Gamma: {\bf q}=(0,0,0)$, X: $(2\pi,0,0)$, 
W: $(\pi,0,2\pi)$,
 L: $(\pi,\pi,\pi)$,
 K: $(\frac{3\pi}{2},\frac{3\pi}{2},0)$, and
 U: $(\frac{\pi}{2},2\pi,\frac{\pi}{2})$, where the lattice constant is 
 set to unity.
}
\label{fig-sw_spectra}

\end{figure}
\begin{figure}[h!]%

\begin{center}
    \includegraphics[width=0.45\textwidth]{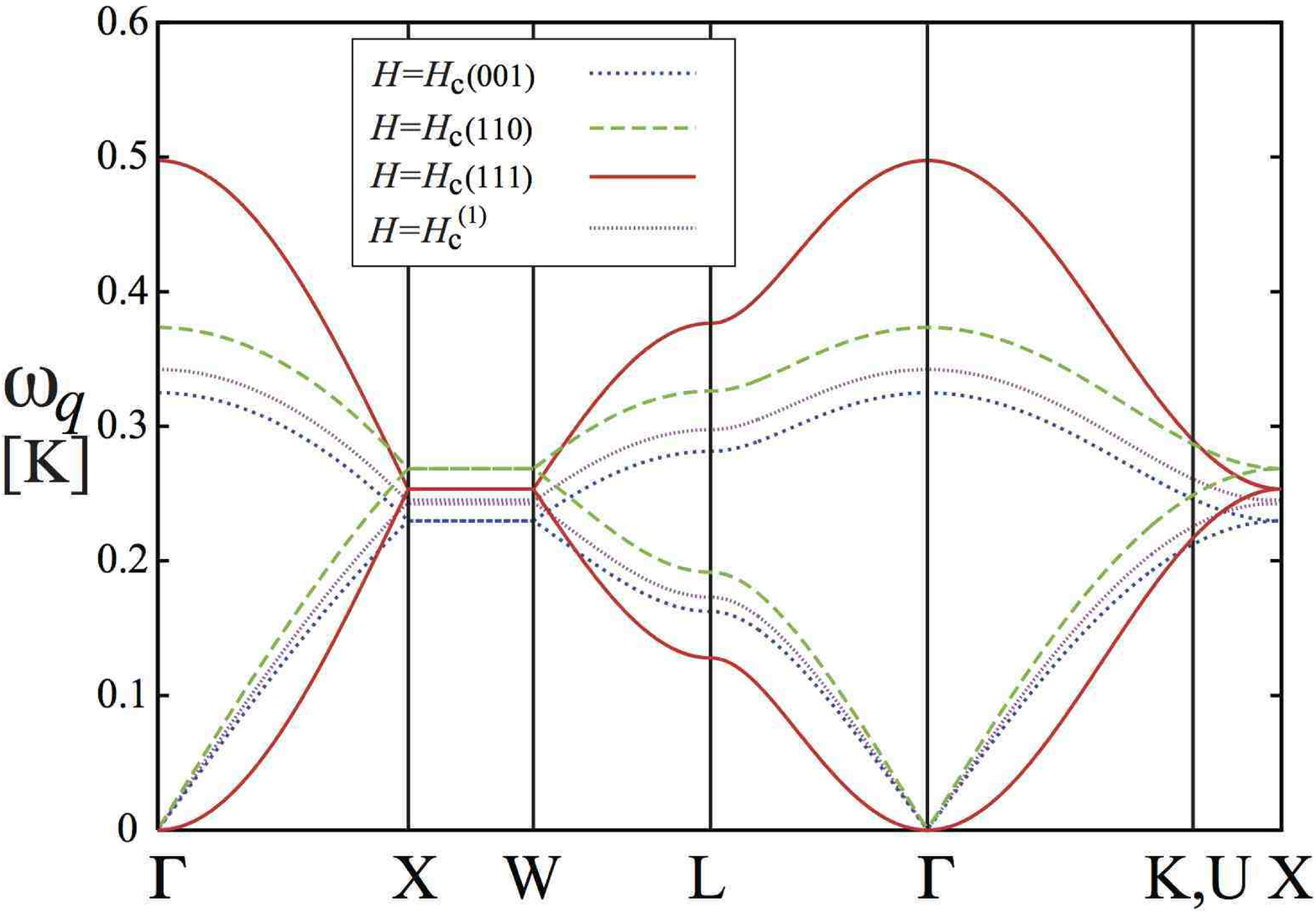}
\end{center}
\caption{(Color online) Excitation spectra at critical fields: $|{\bf
 H}|=H_c(001), H_c(111), H_c(110)$ and $H_c^{(1)}$ for ${\bf
 H} \parallel$ [001].}
\label{fig-sw_spectra2}%

\end{figure}

{
The dependence of the excitation spectra on wavevector ${\bf q}$ is shown
in Fig. \ref{fig-sw_spectra} for ${\bf H} \parallel$ [001]. For other
directions, we obtain similar results, and we do not show them here.
Since the energy scale of the interaction $\lambda$ is much smaller than
local energies in $H_{\rm CEF}$, the high energy modes are similar to 
local CEF excitations and their ${\bf q}$-dependence is very small. Therefore, 
we show here only the lowest
two modes, which consist mainly of the non-Kramers $\Gamma_3$ doublet.
It is noted that the number of the modes does not change in all the
phases, since only ${\bf q}={\bf 0}$ orders are
considered in this paper.

First, we discuss the nature of these two lowest-energy modes.
They are
rotation of ${\bf Q}^{A,B}$ in the two-dimensional ${\bf Q}$-space from
the ground-state configuration in each phase. 
 There are basically two types of rotations.
One is an in-phase rotation of ${\bf Q}^A$ and ${\bf Q}^B$; that is,  
the two quadrupole moments rotate in the same direction. The other
is an out-phase rotation; the two rotate in an opposite direction. We label them by ``in'' and ``out''
 in Fig. \ref{fig-sw_spectra}. Note that in the phase-III
 a level crossing occurs between these two excitations as shown in
Fig. \ref{fig-sw}. This means that ``in'' and ``out'' are interchanged
for the higher
field part of the phase-III. Figure \ref{fig-sw_spectra} (c) shows 
the spectra in the lower field part.  

To understand the nature of each of these modes, it is easiest to study the
high-field phase-IV. In the high-field phase, the
lowest-energy excitation is the ``out'' mode and thus has a finite matrix
element of staggered $Q_x$ with the
ground state. Since 
the system starts to exhibit the $Q_x$ antiferro quadrupole order as the magnetic field decreases, 
this is natural and it is this softening that leads to the
phase transition. For other phases, similar arguments are also possible. 

Secondly, we note that the flat band in the excitation spectrum along
X-W direction has the same energy as the lowest excited level of an
isolated single site calculated by the mean-field theory. This is because
 the form factor of the 
diamond lattice structure vanishes there. One can see that the two modes
 in Fig. \ref{fig-sw_spectra} are degenerate along X-W except for (b). This is related to the symmetry
between $A$- and $B$-sublattices for $|{\bf H}|=0$ and $H_c^{(2)}<|{\bf
H}|$, while there is no such symmetry for $0<|{\bf
H}|<H_c^{(2)}$, since $|{\bf Q}^A|\ne|{\bf Q}^B|$. 

Finally, we show the excitation spectra at the critical fields in Fig. \ref{fig-sw_spectra2}.
At or near the continuous transition for ${\bf H} \parallel$ [001] and [110],
 the energy dispersion is $|\bf q|$-linear around the $\Gamma$ point, 
 which also indicates that the system becomes critical. 
At the critical field for ${\bf H} \parallel$ [111], however, the dispersion is
$|{\bf q}|^2$ and this is due to the absence of induced quadrupole
moment in this field direction. This is qualitatively understood
by considering an effective pseudospin-1/2 XY model with a field
perpendicular to the XY plane. This corresponds to the fact that the
inter-site interactions are only about the quadrupoles, while the magnetic field
couples with the octupole when ${\bf H} \parallel$ [111].
Within the spin-wave approximation,
anomalous terms $a^{\dagger}_ib^{\dagger}_j$ and $a_ib_j$ vanish and only hopping
terms $a^{\dagger}_ib_j$ and $b^{\dagger}_ja_i$ remain in
Eq. (\ref{H2b}). 
This is the same for spin waves in isotropic ferromagnets and
there the dispersion is $|{\bf q}|^2$.
}

\subsection{Quadrupole susceptibilities} \label{Qchi}
In this subsection, we study the quadrupole susceptibilities.
We first show the numerical results of the quadrupole susceptibilities
as a function of magnetic field for three symmetric directions in
Sect. \ref{chiQ}. As we will show, unusual critical behaviors appear for
${\bf H} \parallel$ [111], and to clarify them, we analyze the nature of
the unusual criticality in Sect. \ref{GLsus}.

\subsubsection{Numerical results}\label{chiQ}
\begin{figure}[t]
\begin{center}
    \includegraphics[width=0.5\textwidth]{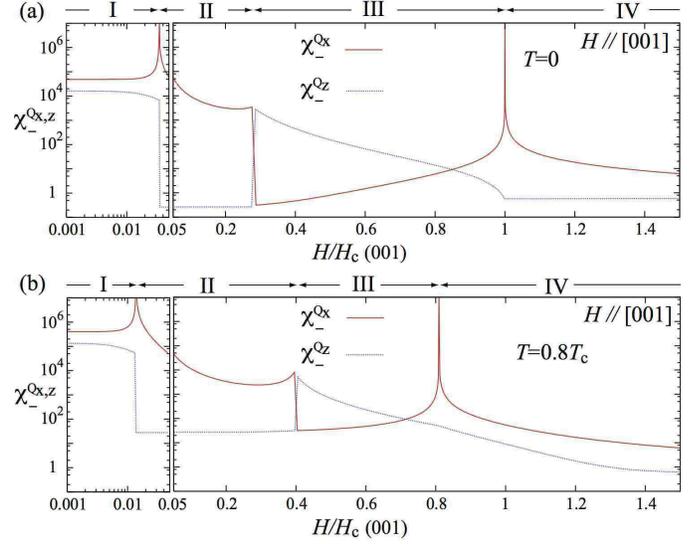}
\end{center}
\caption{(Color online) Quadrupole susceptibility $\chi_-^{Qx,Qz}$ vs. magnetic 
 field ${\bf H} \parallel [001]$. (a) $T=0$ and (b) $T=0.8T_c({\bf H}={\bf 0})$.} 
\label{fig-chiQ}
\end{figure}

Let us first define two components of the
susceptibilities with respect to the sublattice indices. 
Because of $A$ and $B$ sublattices, 
the quadrupole susceptibility is a 2$\times$2 matrix, 
and we consider two parity-conserving responses that are also
diagonal in ${\bf Q}$-space. Just for notational convenience, we denote them in terms of
$\underline{\chi}^{\pm}$ in Eq. (\ref{finalchiMF}) as
$\chi^{Q\mu}_{\pm}
=(\underline{\chi}^{\pm})_{\mu\mu}$,
where $\mu=z$ or $x$.
 In the following, we will study the behavior of these 
quadrupole susceptibilities upon changing 
magnetic field strength for three special 
field directions. 

\begin{itemize}
\item ${\bf H} \parallel$ [001]:
\end{itemize}

Figure \ref{fig-chiQ} shows the staggered part of the static quadrupole susceptibility, 
$\chi_{-}^{Q\mu}$ as a function of magnetic field applied along [001] direction.  
Note that the staggered part here refers to the component 
with $\mathbf{q}=\mathbf{0}$ 
that has an odd parity for exchange of $A$ and $B$ sites, 
instead of finite-$\mathbf{q}$ component.  
The most prominent feature is divergence at two phase boundaries.  
The divergence at higher field takes place 
at the transition between phases III and IV, 
while the divergence at lower field concurs with 
the transition between phases I and II at $H_c^{(1)}\sim 0.03 H_c(001)$.  
The staggered order parameter, 
$\langle Q_x^A \rangle - \langle Q_x^B \rangle$, 
is zero in the phases II and IV, and 
this continuously emerges upon moving into the phases I and III. 
The divergence of $\chi_{-}^{Qx}$ is a consequence of 
this continuous phase transition.  
Another prominent feature is a large jump at $H_c^{(2)}\sim 0.3 H_c(001)$.  
This is due to the first-order phase transition 
between the phases II and III.   
Just above $H_z^{(2)}$ in the phase III, the quadrupole moments 
are aligned almost parallel to $Q_x$ direction, 
and their size is large $|\langle Q_x^{A,B} \rangle | \sim 1$, 
as shown in Fig. \ref{fig-op}(a).  
Therefore, their fluctuations are suppressed and this 
leads to a reduction of $\chi_{-}^{Qx}$.  
The quadrupole order is completely different 
in the phase II, which is below $H_c^{(2)}$.  
The quadrupole moments are aligned to $\pm Q_z$ direction.  
Therefore, their transverse fluctuations contribute to 
$\chi_{-}^{Qx}$, and its value is quite large.
\begin{figure}[t]
\begin{center}
    \includegraphics[width=0.5\textwidth]{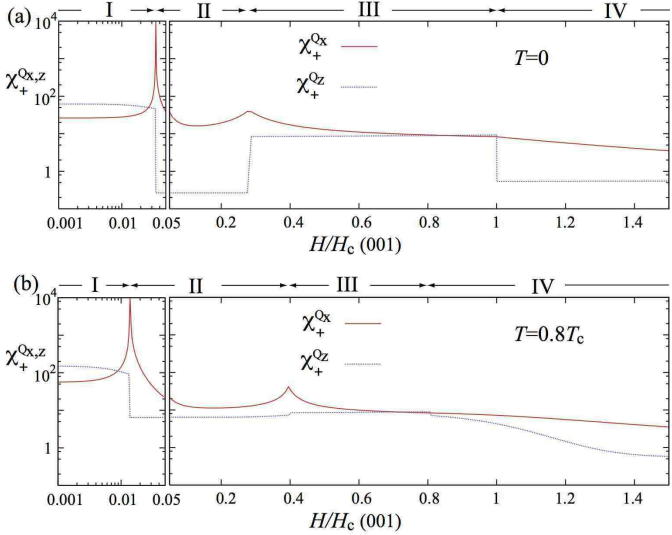}
\end{center}
\caption{(Color online) Quadrupole susceptibility $\chi_+^{Qx,Qz}$ vs. magnetic 
 field ${\bf H} \parallel [001]$ (a) $T=0$ and (b) $T=0.8T_c({\bf H}={\bf
 0})$.}

\label{fig-chiQuni}
\end{figure}

Singularity is found also in the uniform part of 
the static susceptibility, 
$\chi_{+}^{Qx}$, as
shown in Fig. \ref{fig-chiQuni}, 
although this singularity is weaker than that in 
the staggered part $\chi_{-}^{Qx}$.   
At the transition point between the phases IV and III, 
the uniform susceptibility does not diverge but shows a kink. See the
(red) line near $H\sim H_c(001)$. 
The absence of divergence comes from the fact that 
the corresponding static quantity 
$\langle Q_x^A \rangle + \langle Q_x^B \rangle$ is not 
an order parameter in either of the phases III and IV. 
However, this quantity couples to the order parameter 
and this leads to a kink singularity in its field dependence.  
This is similar to the singularity in uniform susceptibility 
of antiferromagnets at the Neel temperature.  
This kink behavior is enhanced with lowering temperature.  
The transition between the phases III and II is first order, 
and $\chi_{+}^{Qx}$ shows a small jump there.  
The transition between the phases II and I is special, 
and the uniform susceptibility also diverges.  
This is because in the phase I 
$|\langle Q_x^A \rangle | \ne  |\langle Q_x^B \rangle |$, while 
both vanish in the phase II.  
Therefore, not only the staggered part but also 
the uniform part have fluctuations that diverge 
with approaching the transition point, which 
leads to $\chi_{\pm}^{Qx} \rightarrow \infty$.  
As for $\chi^{Qz}_+$, it is suppressed in collinear phases with $\langle
Q_x\rangle=0$ such as in the
phases II and IV, while enhanced in the phases I and III. This is
natural, since longitudinal fluctuations are expected weaker than
transverse ones.

\begin{itemize}
\item ${\bf H} \parallel$ [110]:
\end{itemize}

\begin{figure}[t!]

 \begin{center}
    \includegraphics[width=0.45\textwidth]{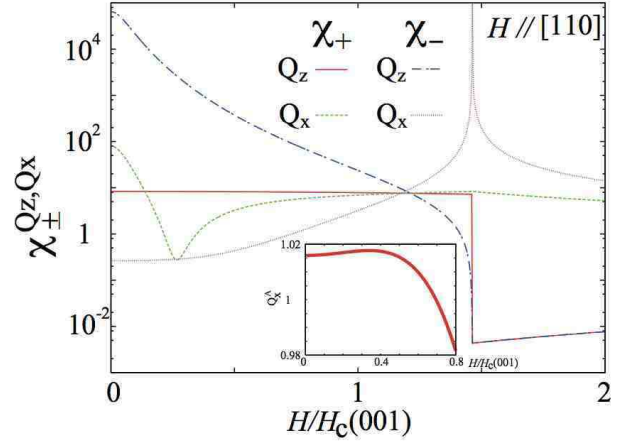}
 \end{center}
 \vspace{-3mm}
 \caption{(Color online) Quadrupole susceptibility vs. 
   magnetic field for  ${\bf H} \parallel [110]$.
 Inset: $\langle Q^A_x\rangle=-\langle
  Q^B_x\rangle$ at low fields.}
 \label{fig-chiQ110-111}
 \vspace{-5mm}
 \end{figure}

Figure \ref{fig-chiQ110-111} shows $\chi^{Qz,x}_{\pm}$ at $T=0$ and for
${\bf H} \parallel [110]$. In addition to
the divergence in $\chi_-^{Qx}$ at the critical field, one can see a
dip in $\chi_+^{Qx}$ at 
low fields. Indeed, $\langle Q_x\rangle$ shows a non-monotonic
behavior in the low-field regime as shown in the inset of
Fig. \ref{fig-chiQ110-111}, and more directly, the weight of
the lowest-energy eigen mode related to $Q_x^A+Q_x^B$
vanishes at the magnetic field where $\chi_+^{Qx}$ is a minimum,
while the change in eigenenergy is monotonic as shown in Fig. \ref{fig-sw}.

\begin{itemize}
\item ${\bf H} \parallel$ [111]:
\end{itemize}

\begin{figure}[t!]

 \begin{center}
    \includegraphics[width=0.4\textwidth]{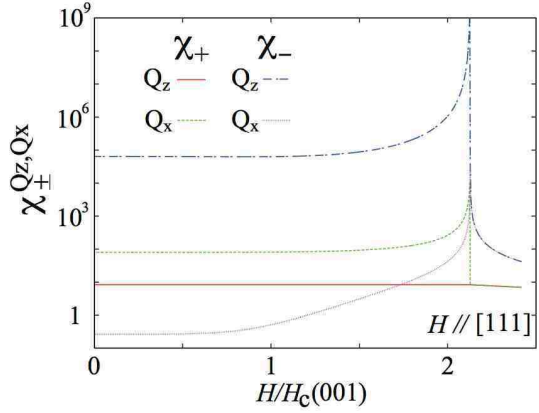}
 \end{center}
 \vspace{-3mm}
 \caption{(Color online) Quadrupole susceptibility vs. magnetic field for ${\bf
  H} \parallel$ [111]. }
 \label{fig-chiQ111}

 \end{figure}

For ${\bf H} \parallel [111]$, there is only one transition and 
the susceptibilities show unusual
magnetic-field dependence near the critical field. 
To be specific, 
 we choose the $(x^2-y^2)$ domain in the ordered phase 
[see Fig. \ref{fig-phaseconfig2} (c)]. 
Figure \ref{fig-chiQ111} shows that not only the staggered 
components $\chi_{-}^{Qz,Qx}$ but also 
 the uniform component $\chi_+^{Qx}$ diverge at the critical field $H_c(111)$.
It is interesting to note that $\chi_+^{Qx}$ diverges  {\it only}
in the ordered phase. The singularities of static $\chi_{\pm}^{Qz,Qx}$ are summarized in Table \ref{tbl-1}.
 Note that the strong singularity for $\chi_-^{Qz}\propto |\delta
H|^{-2}$ in the ordered phase makes a striking contrast to the
conventional mean-field divergence $\sim |\delta H|^{-1}$ in the other
field directions. We will investigate 
the origin for these unusual behaviors in the next subsubsection.


\begin{table}[h!]
\caption{Singularity of the quadrupolar susceptibilities near the
 critical field $H_c(111)$ for ${\bf H} \parallel$ [111] and those near
 the critical temperature $T_c$ at ${\bf H}={\bf 0}$. Here, $\delta H=H-H_c$ and
 $\delta T=T-T_c$.}\vspace{3mm}
\begin{center}
\begin{tabular}{c|cc|cc}
\hline\hline
\multicolumn{1}{c}{\ } & \multicolumn{2}{|c}{${\bf H} \parallel$ [111]} &
\multicolumn{2}{|c}{${\bf H}={\bf 0}$}\\
& $H<H_c$& $H>H_c$ & $T<T_c$ & $T>T_c$\\
\hline
$\chi_-^{Qz}$ & $|\delta H|^{-2}$& $\delta H^{-1}$ & $|\delta T|^{-2}$
	     & $\delta T^{-1}$ \\
$\chi_-^{Qx}$ & $|\delta H|^{-1}$& $\delta H^{-1}$ & $|\delta T|^{-1}$
	     & $\delta T^{-1}$ \\
$\chi_+^{Qz}$ & $\sim$const.& const. $-\delta H$ & $\sim$const.
	     & const. $-\delta T$ \\
$\chi_+^{Qx}$ & $|\delta H|^{-1}$& const. $-\delta H$ & $|\delta T|^{-1}$
	     & const. $-\delta T$ \\
\hline\hline
\end{tabular}
\end{center}
\label{tbl-1}
\end{table}


\subsubsection{Analysis on unusual divergences in susceptibilities } \label{GLsus}
In this subsubsection, we 
show that the $A$-$B$ sublattice mean-field approximation can explain those discovered 
unusual divergences in $\chi$'s if the $Z_3$ anisotropy in ${\bf
Q}$-space is correctly taken into account.
Recall that this occurs for increasing magnetic field ${\bf
H} \parallel$ [111]. As will be discussed, this also happens 
near the transition temperature at ${\bf H}={\bf 0}$.

 In Appendix
\ref{susMF}, we show the formula of the uniform and staggered quadrupole
susceptibilities $\underline{\chi}^{\pm}$ in terms of the local
susceptibility $\underline{\chi}_{\rm loc}^{A,B}$. In this subsubsection, we will
analyze
$\underline{\chi}_{\rm loc}^{A,B}$ calculated by the Landau theory developed
in Appendix \ref{VariLandau}, and show that the susceptibilities indeed
exhibit unusual singularities as discussed in the previous subsubsection.

First, let us calculate the local susceptibility from single-site free
energy $F_{\rm AFQ}^s$ valid for ${\bf H} \parallel$ [111] or ${\bf
H}={\bf 0}$:
\begin{eqnarray}
F_{\rm AFQ}^{s}\!\!\!\!\!&=&\!\!\!\!\! \frac{a}{2}|{\bf
 Q}^s|^2-\frac{\gamma}{3}[Q^s_z({Q^s_z}^2-3{Q^s_x}^2)]
+\frac{b}{4}|{\bf Q}^s|^4,\label{Fsite}
\end{eqnarray}
where $s=A$ or $B$ is the site index.
We consider a domain in  the ordered phase where $\langle Q_x^A\rangle=-\langle
Q_x^B\rangle=q_s$ and $\langle Q_z^{A,B}\rangle=-q_u$. Here, $q_s$ and
$q_u$ are given by Eqs. (\ref{eq-G:Rmin}) and (\ref{qvalue}).  
The free energy $F^s_{\rm AFQ}$ is expanded up to the second order in the
fluctuation $\delta {\bf Q}^s={\bf Q}^s-\langle {\bf Q}^s \rangle$ and we obtain
\begin{eqnarray}
F^s_{\rm AFQ}=F^{s0}_{\rm AFQ}+\frac{1}{2}\delta {\bf Q}^s \cdot \Big(\underline{\chi}^s_{\rm loc}\Big)^{-1} \cdot \delta
 {\bf Q}^s.
\end{eqnarray}
Here, $F_{\rm AFQ}^{s0}$ is the free energy at the stationary point. 
In this domain, $\underline{\chi}_{\rm loc}^B$ 
is identical to $\underline{\chi}_{\rm loc}^A$  except the sign
of the off-diagonal elements. 
In terms of the parameters in Appendix \ref{VariLandau}, we
obtain 
\begin{eqnarray}
\Big(\underline{\chi}_{\rm loc}^A\Big)_{\mu\mu} &\sim& 
\frac{1}{g}+c_{\mu}^{(1)}\delta h +c_{\mu}^{(2)} \delta h^2,\\
\Big(\underline{\chi}_{\rm loc}^A\Big)_{zx} &\sim& 
t_1\delta h^{1/2} +t_2 \delta h^{3/2},
\end{eqnarray}
where $g$ is the effective intersite quadrupole coupling and $\delta
h=g-a$ is the control parameter representing the distance from the
critical point, and 
\begin{eqnarray}
c_z^{(1)}\!\!\!\!&=&\!\!\!\!\frac{2\gamma^2/g}{g^2\tilde{b}}, \ 
c_x^{(1)}=-\frac{2(b-2\gamma^2/g)}{g^2\tilde{b}}, \ 
t_1=\frac{2\gamma}{g^2\sqrt{\tilde{b}}},\ \ \ \ 
\end{eqnarray}
with $\tilde{b}=b-\gamma^2/g$.
Pay attention to the relation
\begin{eqnarray}
c_z^{(1)}=\frac{1}{2}gt_1^2,
\end{eqnarray}
and this is the {\it key} to the unusual singularities. 
This results in a cancellation of the $\delta
h^1$-order term in the calculation of
det$(1-g^2\underline{\chi}_{\rm loc}^A\underline{\chi}_{\rm loc}^B)$
 and enhances singularities in various channels. Calculating the
 higher-order correction coefficients $c_z^{(2)}$, $c_x^{(2)}$, and $t_2$,
we obtain the leading singularities as 
\begin{eqnarray}
\chi_-^{Qz} &\sim&
\frac{4\tilde{b}^2g^3}{3\gamma^2(3\tilde{b}g+2\gamma^2)} \delta h^{-2},\\
\chi_-^{Qx} &\sim& \delta h^{-1},\\
\chi_+^{Qz} &\sim&
\frac{b}{\tilde{b}g},\\
\chi_+^{Qx} &\sim&
\frac{4\tilde{b}g}{3(3\tilde{b}g+2\gamma^2)} \delta h^{-1}.
\end{eqnarray}
These are exactly what we have obtained in Sect. \ref{chiQ}.

\begin{figure}[t!]

 \begin{center}

    \includegraphics[width=0.43\textwidth]{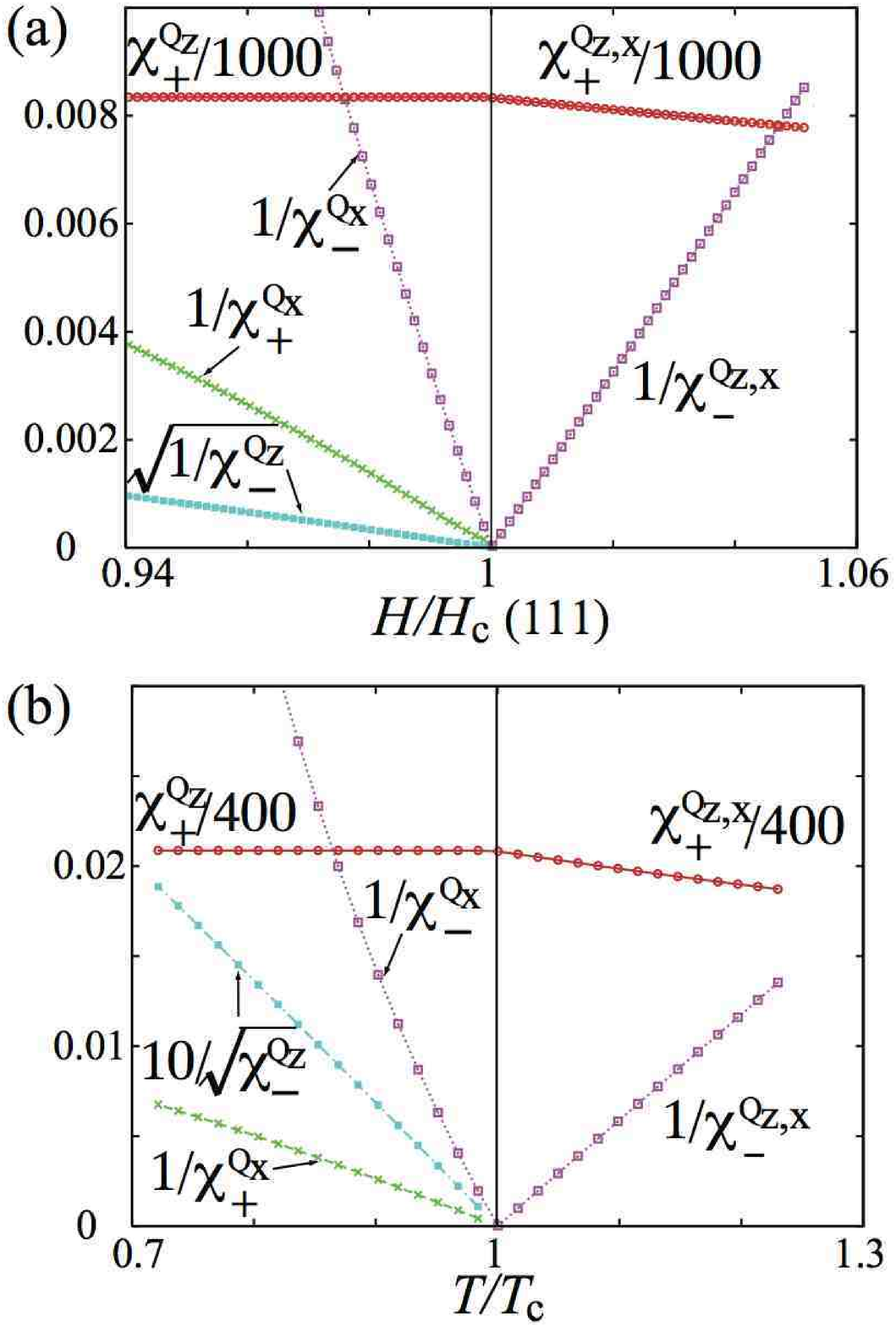}
 \end{center}
 \caption{(Color online) Quadrupole susceptibility near critical points. (a) ${\bf
  H} \parallel$ [111] at $T=0$. (b) ${\bf H}={\bf 0}$.}
 \label{fig-check}. 
\vspace{-5mm}
 \end{figure}

Note that the change of control parameter $g-a$ can be also driven by
temperature, {\it i.e.,} $\delta h = g-a(H,T)$. Therefore, the same
singularities appears 
at $|{\bf H}|=H_c(111)$ for
${\bf H} \parallel$ [111] 
and 
at the critical temperature for ${\bf H}={\bf 0}$ 
by regarding 
$\delta h\propto \delta H$ or $\propto \delta T$ as shown in Table \ref{tbl-1}.
Indeed, we have checked this by microscopic mean-field calculations as
shown in Fig. \ref{fig-check}. 
The result of $\chi_{-}^{Qz}$ and $\chi_{+}^{Qx}$ 
are very unique and interesting. 
When the anisotropy is decreased $\gamma \rightarrow 0$, 
the amplitude of the divergence of $\chi_{-}^{Qz}$ 
itself diverges as $1/\gamma^2$.  
In isotropic systems, we know that the transverse susceptibility 
should be infinite in the whole region of its ordered phase as a
consequence of gapless Goldstone mode. 
The behavior found here is in this sense consistent.

%


\section{Discussions}\label{Discussions}
In this section, we will discuss several topics relevant to Pr-based 1-2-20
compounds based on the results in this paper.

\subsection{Ferro quadrupole order}\label{FQ}
In this paper, we have studied antiferro quadrupole orders, which appear
in several Pr 1-2-20 compounds. As for PrTi$_2$Al$_{20}$, it is suggested that the ordered state is ferro
quadrupole.\cite{Sakai1,Sato,Nakanishi} We briefly discuss below properties of
$\Gamma_3$ ferro quadrupole order.

Also for the ferro quadrupole case, the
discussion on $O_2^2$($Q_x$) order in Sect. \ref{singlesite} holds and it is generally
accompanied by a finite $O_2^0$($Q_z$) component. Neutron
scattering experiment suggests that the order is ferro
$O_2^0$,\cite{Sato} and, in this case,
there is no induced $O_2^2$ moment from our discussions in Sect. \ref{singlesite}.

An important difference from the antiferro quadrupole order is the order
of the transition. It is generally first-order as predicted by  the
Landau theory. The Landau free energy
$F_{\rm FQ}$ is given as 
\begin{eqnarray}
 F_{\rm FQ} \sim \frac{1}{2}a |{\bf Q}|^2+\frac{1}{3}\gamma Q_z(Q_z^2-3Q_x^2)
  + \frac{1}{4}b |{\bf Q}|^4. \label{FQF}
\end{eqnarray}
Here, $a$, $b$, and $\gamma$ are constants and  ${\bf Q}$ is the uniform quadrupole moment. Note that for
antiferro quadrupole case ${\bf Q}_{\rm AF}$, the third-order term
including three ${\bf Q}_{\rm AF}$'s is not present due to 
the inversion symmetry ($A\leftrightarrow B$). The possibility of the
first-order transition for the antiferro case is discussed in Appendix \ref{VariLandau}.

The third-order anisotropy $\gamma$ is generally finite unless
microscopic parameters are finely tuned.
 Thus, the transition should be
first-order in general cases.\cite{LandauLifshitz} The order of the transition can be
controlled by, for example, applying magnetic field. Since the quadrupole couples with
magnetic field through Eq. (\ref{Hmag2}), there appears a linear term in
$F_{\rm FQ}$, leading to a finite moment $\langle \bf Q\rangle$
induced. Then, the situation is similar to the classical liquid-gas
transition and 
it is possible to tune the
system to a second-order transition point and also crossover regime by
varying the magnetic field and temperature. 

Recently, Matsubayashi et
al., observed that the superconducting transition temperature for
PrTi$_2$Al$_{20}$ is enhanced near the region where the ferro quadrupole
order disappears under pressure.\cite{Matsubayashi} 
We expect that this is due to critical or strongly enhanced quadrupole
fluctuations of orbital degrees of freedom.  However, if
the order is ferro quadrupole, the transition is generally
first order as discussed before.  Then, fluctuations are
not so particularly enhanced near the transition, unless
the transition is very weak first order.  For quantitative
comparison with experimental data, we need more elaborate
calculations with parameter tuning, which is left for a
future study.

\subsection{Softening in elastic constants in PrIr$_2$Zn$_{20}$}
Uniform quadrupole susceptibility is measured indirectly by ultrasonic
experiments
 through the coupling of elastic constant and quadrupole susceptibility.\cite{Levy}
In systems with $T_d$ symmetry, one measures the elastic constants 
$c_{11}-c_{12}$ and the $c_{44}$ 
to detect $\Gamma_3$ and $\Gamma_5$ quadrupoles, respectively.
$c_{11}-c_{12}$ gives information about uniform quadrupole
susceptibilities in $\Gamma_3$ sector:
$\chi^{Qz,Qx}_+$, while $c_{44}$ does about $\Gamma_5$
quadrupoles, which are due to the excited states in the 1-2-20 compounds. 
 In PrIr$_2$Zn$_{20}$, the elastic constants
exhibit unusual behaviors as a function of magnetic field and
temperature.\cite{Ishii1} In this subsection, we discuss two aspects
of them.

For ${\bf H}={\bf 0}$, the elastic constants
exhibit softening near the transition temperature 
both in the $c_{11}-c_{12}$ and the $c_{44}$
modes.
Since the CEF ground state is the $\Gamma_3$ doublet, the softening in
$c_{11}-c_{12}$ is due to its degeneracy, which leads
to $\sim -1/T$ at low temperature for $T>T_c$ (note that this does not
diverge at $T_c$, since the transition occurs in the  antiferro
quadrupole sector).
However, that in $c_{44}$ mode cannot be explained by a simple picture,
since the excited states are in high energy above $\sim$ 30 K and there is no $\Gamma_5$ in the direct product
$\Gamma_3\otimes\Gamma_3$. See, Appendix \ref{direct}. 
As a source of the softening in
the $c_{44}$ mode, the effects of mode-mode coupling might be important in
this compound. For $c_{11}-c_{12}$ mode, our result in Sects. \ref{chiQ}
and \ref{GLsus} demonstrates that one of $\chi^{Qx}_+$ and $\chi^{Qz}_+$
susceptibilities depending on its domain diverges in the ordered phase toward
the critical temperature and this is consistent with the experiment.

Under magnetic fields, the elastic constants also
exhibit softening as a function of the magnetic field near the
high-field critical point.\cite{Ishii1} 
When the magnetic field is applied in [100] direction, 
 $c_{11}$
( a part of $\Gamma_3$ mode) 
shows strong softening as a function of magnetic field and also of
temperature at 5 T. 
 In addition, there are two anomalies below 5 T, suggesting
the existence of multiple phases under the magnetic field. 

For other directions, the elastic constant
in $\Gamma_3$ mode also shows softening.
Our mean-field result is consistent with those for ${\bf
H} \parallel$ [111] in the ordered phase, while in other cases the origin
of the softening is beyond the mean-field approximation. In addition,
there are several anomalies for ${\bf H} \parallel$ [110] and
[111].\cite{Ishii1,Ishii2} Further experimental works will clarify the
nature of the anomalies and the whole $T$-$H$ phase diagram.

It is noted that the number of phases for ${\bf H} \parallel$ [001] suggested in these experiments is consistent
with the present results, which is also consistent with the early
analysis by Onimaru.\cite{OnimaruP2} 
Thus, one can expect that the
strong softening around 5 T is related to the quantum critical point of 
antiferro quadrupole order between the phase-III and the phase-IV.
This is a promising scenario and we need to carry out more elaborate
calculations beyond the mean-field approximation, since
 the diverging susceptibility is not the uniform quadrupole susceptibility but the
 antiferro quadrupole one at $H_c(001)$ for ${\bf H} \parallel$ [001]. 

It is also instructive to point out that the uniform quadrupole
susceptibility diverges when the sizes of corresponding quadrupole
moments at 
two sublattices are different and they vanish at the transition. 
This situation is realized between the phase-I and the phase-II.
The origin of the enhancement in the uniform
quadrupole susceptibility near the transition between 
the phase-II and the phase-III is also related to this.

\subsection{Thermo-electric power}
Recently, Izawa {\it et al}., observed strong enhancement in thermo-electric
power $S$ in PrIr$_2$Zn$_{20}$ as a function of temperature under high
magnetic fields.\cite{Izawa} The peak position coincides with the peak
position of specific heat.\cite{OnimaruP} With decreasing magnetic field, the peak position shifts to
a lower temperature and seems to vanish at the critical field $\sim 5$ T. 
 This suggests that the peak position is related to
some energy scale of dynamics.  A candidate of this energy scale is the 
``spin''-wave gap at ${\bf q}={\bf 0}$ as discussed in Sect. \ref{SWA2}.
This explains the ${\bf H}$ dependence of specific heat peak position
and the fact
 that the peak 
temperature vanishes at the critical field. 
Since in Kondo
systems,  a peak appears around the Kondo temperature,\cite{Lac,Zlatic}
 to clarify which energy scale determines  the
peak in $S$, one needs to carry out more elaborate calculations 
 for non-magnetic $\Gamma_3$ systems including both the on-site  Kondo screening and
 inter-site correlations.

\subsection{Other intersite interactions} \label{otherint}
In the present study, we have studied the canonical and minimal model
with only quadrupole coupling, and have not included magnetic interactions or 
other non-magnetic ones. 
To quantitatively reproduce
 the phase diagram observed in the experiments,
it is necessary to include other
interaction, {\it e.g}., magnetic dipole interaction and also to carry
out calculations beyond the mean-field approximation. 
Here, we examine the types of possible interactions based on symmetry
arguments. See, details in Appendix \ref{how}.

In Pr-based 1-2-20 compounds, Pr ions form a
diamond lattice structure. In the following, we list some of the
nearest-neighbor interactions possible in this case.

First, as for the quadrupole-quadrupole interactions, only  one type is 
possible, and this is the one used  in the present paper.

Second, as for the dipole-dipole interactions, there are two types allowed and
they are given as
\begin{eqnarray}
g_1{\bf J}(i)\cdot {\bf J}(j)+
g_2\big[\hat{r}_{ji}\cdot {\bf J}(i)\big]
\big[\hat{r}_{ji}\cdot {\bf J}(j)\big].\label{KH}
\end{eqnarray}
Here, $i$ and $j$ are nearest neighbors on different sublattices 
and 
$\hat{r}_{ji}={\bf r}_{ji}/|{\bf r}_{ji}|$ is
the unit vector from $i$ to $j$. 
The first term is isotropic in both of ${\bf J}$ space and real space,
while the second is anisotropic in both spaces.

Thirdly, as for the octupole moments, in addition to trivial $T_{xyz}(i)T_{xyz}(j)$ type interaction, the
$T_{xyz}$ octupole can couple with dipole moments as
\begin{eqnarray}
g_3\Big\{T_{xyz}(i) \big[\hat{r}_{ji} \cdot {\bf J}(j)\big]+T_{xyz}(j) \big[\hat{r}_{ji} \cdot {\bf J}(i)\big]\Big\}.
\end{eqnarray}
There are many others, but we stop here and leave them in future publications.

It is noted that the interactions listed above include directional ones,
{\it i.e.,} those including $\hat{r}_{ji}$. These anisotropic interactions
are in general important in the f-electron systems and might lead to
different ground states from the simple $\Gamma_3$-$\Gamma_3$ model. 
In order to determine these coupling constants, inelastic neutron
scattering is powerful and one can compare the spin-wave dispersions
between the experiments and the theory.

\subsection{Interactions with phonons} \label{otherint2}
\begin{figure}[t!]

\begin{center}
    \includegraphics[width=0.4\textwidth]{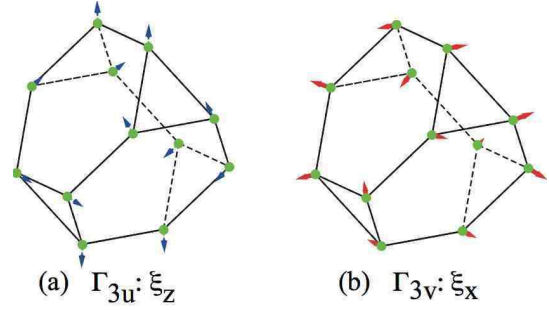}
\end{center}
\caption{(Color online) An example of local $\Gamma_3$ phonon mode of the cage
 X$_{12}$ (X=Zn, or Al). Arrows show directions of displacement. (a) $\Gamma_{3u}:\xi_z$ and (b) $\Gamma_{3v}:\xi_x$.}
\label{fig-cageEg}

\end{figure}

Quadrupole moments are located at the center of a cage and they couple with local
phonons of the cage atoms. In particular for the $\Gamma_3$-type mode of
displacements 
denoted as $\vec{\xi}(i)=[\xi_z(i),\xi_x(i)]$ (see
Fig. \ref{fig-cageEg}), 
this couples linearly with the quadrupole at the center  as 
$
 {\bf Q}(i)\cdot \vec{\xi}(i)
$.

\begin{figure}[t!]

\begin{center}
    \includegraphics[width=0.2\textwidth]{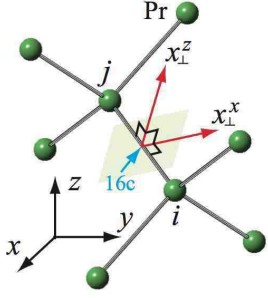}
\end{center}
\caption{(Color online) Transverse displacement ${\bf x}_{\perp}(ij)$
 at the 16c site located at the bond center. }
\label{fig-xperp}

\end{figure}

It would be more interesting to consider 16c site displacements in
the 1-2-20 compounds. 
The 16c site (Zn or Al) is located at the middle of a nearest-neighbor Pr-Pr bond.
The first-principle calculations\cite{HasegawaLDA} for La compounds
suggest 
that its atom oscillation  is
highly anharmonic and anisotropic with the hard axis along the
bond direction. We denote the displacement of this oscillation at the
center of the $ij$ bond as ${\bf x}_{ij}\equiv (x_{ij},y_{ij},z_{ij})$.
It can couple with the quadrupole pair as
\begin{eqnarray}
g_4[{\bf Q}(i)-{\bf Q}(j)]\cdot {\bf x}_{\perp}(ij),
\end{eqnarray}
where ${\bf x}_{\perp}(ij)=[x_{\perp}^z(ij),x_{\perp}^x(ij)]\equiv [
(2r^z_{ji}z_{ij}-r^x_{ji}x_{ij}-r^y_{ji}y_{ij})/\sqrt{6},
(r^x_{ji}x_{ij}-r^y_{ji}y_{ij})/\sqrt{2}]
$ is the transverse component perpendicular to the bond
direction as shown in Fig. \ref{fig-xperp}. This evidently indicates that the presence of antiferro quadrupole
order induces a static displacement ${\bf x}_{\perp}(ij)$ in the plane perpendicular to
the bond direction. The direction in the plane is determined by the type
of the quadrupole order. This also causes inversion symmetry breaking,
since the 16c site is an inversion center.

Concerning the displacement ${\bf x}_{\perp}(ij)$, it is also
 interesting that this induces 
 the Dzyaloshinskii-Moriya interactions. This is given by 
\begin{eqnarray}
{\bf D}_{ij}\cdot \big[ {\bf J}(i)\times {\bf J}(j)\big],\label{DMdipole2}
\end{eqnarray}
where the Dzyaloshinskii-Moriya vector is 
${\bf D}_{ij}\propto {\bf x}_{\perp}(ij)\times \hat{r}_{ji}$ 
in the lowest order in the displacement. 
This term favors incommensurate magnetic ordered states in general.
There are many other interactions induced by ${\bf x}_{\perp}(ij)$, but we leave
them for  future studies.

\section{Summary} \label{Summary}
We have investigated antiferro quadrupole orders in
the systems where local ground state is a 
non-Kramers $\Gamma_3$ doublet.
We have considered this system on a diamond lattice to discuss orders in
Pr-based 1-2-20 compounds, but most of the results in the present study
also hold for general bipartite lattices with cubic symmetry. 
We have analyzed a minimal model for antiferro quadrupole orders based
on the mean-field approximation and the quantum fluctuations are also
analyzed via ``spin''-wave calculations.

We have clarified how anisotropy in crystalline-electric-field potential
affects the quadrupole order parameter ${\bf Q}$, and thus, the phase
diagrams under magnetic fields. The third-order nontrivial coupling in
the quadrupoles is essential to explain it. One important consequence
 is that there is no pure $O_2^2$ antiferro quadrupole order
without fine tuning of control parameters and the $O_2^2$ antiferro quadrupole
order is accompanied by ferro $O_2^0$ quadrupole moments. 

The magnetic field-temperature phase diagram shows varieties of phases.
They are explained by competitions between the nonlinear Zeeman coupling 
 (\ref{Hmag2}) and antiferro quadrupole inter-site couping $\lambda$.
For ${\bf H} \parallel [001]$, three ordered phases appear apart from the 
high-field phase where the quadrupole moments align as determined by the
nonlinear Zeeman coupling. The first is a
low-field $O_2^2$ antiferro quadrupole phase. The second is a collinear
$O_2^0$ antiferro quadrupole state. The third is a canted phase. 
For ${\bf H} \parallel [110]$ and [111], there is only one ordered phase and
it is a  canted state.   

To examine excitation dynamics of quadrupole moments, we have analyzed
excitation spectra by using ``spin''-wave approximation. At the phase
boundaries of the second-order transition, there appears a critical mode
with linear energy dispersion $\omega \sim |{\bf q}|$ at the $\Gamma$ point.

We have discovered unusual singularities at $T_c$ for ${\bf H}={\bf 0}$ and
 also at the critical magnetic field along [111] direction, $H_c(111)$, and
clarified their origin.
One staggered quadrupole susceptibility shows a stronger divergence than
conventional mean-field one, and more interestingly, one uniform
quadrupole susceptibility also exhibits divergence, although the ordered
phase is not ferro quadrupole one. This can partially explain the
softening in the ultrasonic experiments, but for the complete
understanding, we need further investigations.

We have also proposed that a possible origin of enhanced thermo-electric power 
 in the high-field phase is related to this low-energy 
quadrupole excitation. This scenario is consistent with the fact that 
 the position of Schottky peak of the
 specific heat $C(T)$ at high fields roughly coincides the peak position
 in the thermo-electric power.

As for possible inter-site couplings in the 1-2-20
compounds, we have shown that there exist various directional
couplings. 
Important ones are the
dipole-dipole interactions and dipole-octupole interactions. 
Quadrupole-phonon interactions are also presented and we have shown that
the atoms at
the 16c site should displace in a way determined by the quadrupole order
pattern. This indicates that if the 1-2-20 system shows an antiferro
quadrupole order, the inversion symmetry of the lattice structure should
be broken at the same time, 
 and as a result, the Dzyaloshinskii-Moriya
interactions are induced. We have shown the form of the
Dzyaloshinskii-Moriya vector in terms of the displacement at the 16c site.
In the present paper, we have concentrated on analyzing a simple
quadrupole-quadrupole model. For more quantitative comparison with 
experimental data, these interactions would be important.

In summary, we have investigated antiferro quadrupole order in
$\Gamma_3$ non-Kramers doublet systems. We have pointed out that there
is no pure $O_2^2$ antiferro quadrupole order from general arguments and
the mean-field phase diagram and
excitation spectra have been demonstrated. We have also discovered
 unusual singularities at $T=T_c$ for ${\bf
H}={\bf 0}$  and also the critical field along [111] direction.
In the final part, we have presented a list of some
 nontrivial interactions in the 1-2-20 compounds, which would be 
important for further theoretical and experimental investigations.

\section*{Acknowledgement}
The authors would thank K. Izawa, T. Onimaru, and T. Sakakibara for
fruitful discussions. K. H. thanks H. Kusunose for his valuable comment
on the CEF Hamiltonian. This work was supported by a Grant-in-Aid for
Scientific Research (No. 30456199) from the Japan Society for the Promotion of Science.

\appendix

\section{Wavefunctions} \label{Wavefunc}
In this Appendix, we list single-site wavefunctions classified by the irreducible
representations in the $T_d$ group used in this paper.

First, let us introduce notations. 
For Pr ions, a main configuration of f-electrons is the $J=4$ multiplet and 
its nine levels split as 
$\Gamma_1\oplus\Gamma_3\oplus\Gamma_4\oplus\Gamma_5$ due to $H_{\rm
CEF}$. Two components of $\Gamma_3$ doublet are denoted by
$\{u,v\}$ and components of $\Gamma_4$ and $\Gamma_5$ triplets are
written as $\{\tilde{X},\tilde{Y},\tilde{Z}\}$ and $\{X,Y,Z\}$, respectively.
Representative forms for the irreducible representations 
in terms of spatial coordinates $(x,y,z)$ 
are given as
\begin{eqnarray}
\!\!\!\!\!\!\!\!\!\!&&\!\!\!\! u\sim 3z^2-r^2, \quad v\sim\sqrt{3}(x^2-y^2),\\
\!\!\!\!\!\!\!\!\!\!&&\!\!\!\! \tilde{X}\sim x(y^2-z^2), \ \tilde{Y}\sim y(z^2-x^2), \
 \tilde{Z}\sim z(x^2-y^2),\ \ \ \ \ \ \ \ \\
\!\!\!\!\!\!\!\!\!\!&&\!\!\!\! X\sim x,\quad Y\sim y, \quad Z\sim z,
\end{eqnarray}
where $r^2=x^2+y^2+z^2$. 
Note that, in $T_d$ symmetry, $xyz$ belongs to $\Gamma_1$, and thus,
$x\sim yz$, $y\sim zx$, and $z\sim xy$. The spatial coordinate vector $(x,y,z)$
is $\Gamma_5$, while an angular momentum ${\bf J}$ belongs to $\Gamma_4$.

Using the above notations, the wavefunctions are given as 
\begin{eqnarray}
|\Gamma_1\rangle \!\!\!&=&\!\!\! \frac{1}{\sqrt{12}}\Big[
\sqrt{\frac{5}{2}}(|4\rangle+|-4\rangle)+\sqrt{7}|0\rangle
\Big],\\
|\Gamma_3u\rangle \!\!\!&=&\!\!\! \frac{1}{\sqrt{12}}\Big[
\sqrt{\frac{7}{2}}(|4\rangle+|-4\rangle)-\sqrt{5}|0\rangle
\Big],\\
|\Gamma_3v\rangle \!\!\!&=&\!\!\! \frac{1}{\sqrt{2}}
(|2\rangle+|-2\rangle),\\
|\Gamma_4\tilde{X}\rangle \!\!\!&=&\!\!\! \frac{1}{4i}\Big[
\sqrt{7}(|1\rangle+|-1\rangle)
+(|3\rangle+|-3\rangle)\Big],\\
|\Gamma_4\tilde{Y}\rangle \!\!\!&=&\!\!\! -\frac{1}{4}\Big[
\sqrt{7}(|1\rangle-|-1\rangle)
-(|3\rangle-|-3\rangle)\Big],\ \ \ \\
|\Gamma_4\tilde{Z}\rangle \!\!\!&=&\!\!\! \frac{1}{\sqrt{2}i}
(|4\rangle-|-4\rangle),\\
|\Gamma_5{X}\rangle \!\!\!&=&\!\!\! \frac{1}{4i}\Big[
(|1\rangle+|-1\rangle)
-\sqrt{7}(|3\rangle+|-3\rangle)\Big],\\
|\Gamma_5{Y}\rangle \!\!\!&=&\!\!\! \frac{1}{4}\Big[
(|1\rangle-|-1\rangle)
+\sqrt{7}(|3\rangle-|-3\rangle)\Big],\\
|\Gamma_5{Z}\rangle \!\!\!&=&\!\!\! \frac{1}{\sqrt{2}i}
(|2\rangle-|-2\rangle).
\end{eqnarray}
Here, $|J_z\rangle$ is state with the z-component of the angular
momentum $J_z$ and  $J=4$.

\section{Crystalline electric field Hamiltonian} \label{RelationHcef}
Here, we comment on two representations of the local CEF Hamiltonian. 
The conventional representation is in terms of Stevens operators and it
reads for $J=4$ ion in CEF of $T_d$ symmetry as 
\begin{eqnarray}
H_{\rm CEF}&=&\sum_i\Big\{B_{4}^0[O_{4}^0(i)+5O_{4}^4(i)]\nonumber\\
     &&\ \ \ \ +B_{6}^0[O_{6}^0(i)-21O_{6}^4(i)]\Big\}\label{eqHcef0}
\end{eqnarray}
Here, $O_{n}^m$'s are the Stevens operators and $B_{n}^m$'s are
constants.\cite{Stev}
This form has been widely used to investigate CEF states in f-electron
systems, but we have found that this is equivalent with the one 
represented by only two operators, and they are quadrupole moments
defined as $Q_x=\sqrt{3}O_2^2/8$ and $Q_z=O_2^0/8$. 
This new representation is Eq. (\ref{eqHcefQ}) shown in Sect. \ref{CEF}.
Equation (\ref{eqHcefQ}) explicitly shows the two invariances of $\Gamma_3$ quadrupole
operators we consider. Apart from a trivial constant,
Eq. (\ref{eqHcef0}) reduces to Eq. (\ref{eqHcefQ}) and the 
parameters are related as
\begin{eqnarray}
\epsilon_2=\frac{640}{3}(B_4^0-126B_{6}^0), \quad \epsilon_3=
 -\frac{78848}{9}B_6^0. 
\end{eqnarray}

\section{Mean-field ground-state energy for ${\bf H}={\bf 0}$} \label{app-T0E}
In this Appendix, we discuss the ground-state energy for ${\bf H}={\bf 0}$ on the
basis of two-sublattice mean-field approximation. 
Since there are no matrix elements in the quadrupole operators 
between $\Gamma_{1,3}$ and $\Gamma_{4,5}$ states, the relevant Hilbert space
here is spanned by $\Gamma_3$ and $\Gamma_1$ in our analysis.

First, let us show matrix forms of quadrupole operators in basis
$\{|\Gamma_1\rangle,|\Gamma_3 u\rangle,|\Gamma_3 v\rangle\}$:
\begin{eqnarray}
Q_z\!\!\!\!&=&\!\!\!\!
\begin{pmatrix}
0 & a & 0 \\

a & 1 & 0 \\

0 & 0 & -1 \\

\end{pmatrix}, \quad
Q_x=
\begin{pmatrix}
0 & 0 & a \\

0 & 0 & -1 \\

a & -1 & 0 \\

\end{pmatrix},
\end{eqnarray}
where $a=\sqrt{35}/2$. The basis wavefunctions are explained in
Appendix \ref{Wavefunc}.

We approximate the intersite interactions by a
mean field of quadrupole ${\bf h}=-z\lambda\langle {\bf Q}\rangle=-{\bf q}$. 
Here, $\langle {\bf Q}\rangle$ is the thermal average on the nearest
neighbor sites and $z$ is the number of the nearest-neighbor sites.
Thus, the mean-field Hamiltonian at one site is written as 
\begin{eqnarray}
H_{\rm 1 site}\!\!\!\!&=&\!\!\!\!
\begin{pmatrix}
E_1 & aq_z & aq_x \\

aq_z & q_z & -q_x \\

aq_x & -q_x & -q_z \\

\end{pmatrix},
\end{eqnarray}
where $(q_z,q_x)=z\lambda(\langle Q_z\rangle, \langle Q_x\rangle)=q(\cos
\theta,\sin \theta)$.
The angle parameter is $0 \le \theta < 2\pi$ and $q=\sqrt{q_z^2+q_x^2}$.
Let us first diagonalize the 
$\Gamma_3$ sector. This is done by using the following new basis:
\begin{eqnarray} 
|\Gamma_3 +\rangle = \cos\frac{\theta}{2}|\Gamma_3 u\rangle
-\sin\frac{\theta}{2}|\Gamma_3 v\rangle, \\
|\Gamma_3 -\rangle = \sin\frac{\theta}{2}|\Gamma_3 u\rangle +\cos\frac{\theta}{2}|\Gamma_3 v\rangle.
\end{eqnarray}
Then, in terms of $\{|\Gamma_1\rangle,|\Gamma_3 +\rangle,|\Gamma_3 -\rangle\}$, $H_{\rm
1 site}$ reads as 
\begin{eqnarray}
H_{\rm 1 site}\!\!\!\!&=&\!\!\!\!
\begin{pmatrix}
E_1 & aq\cos\frac{3\theta}{2} & aq\sin\frac{3\theta}{2} \\
aq\cos\frac{3\theta}{2} & q & 0\\
aq\sin\frac{3\theta}{2} & 0 & -q\\
\end{pmatrix}.
\end{eqnarray}
We are interested in situation where $E_1 \gg z\lambda |\langle {\bf Q}\rangle|=q$ and obtain the
ground-state energy
of $H_{\rm 1 site}$ perturbatively as a series of the small
parameter $q/E_1$,
\begin{eqnarray}
\frac{E_{\rm 1 site}^{\rm gs}}{E_1} \!\!\!\!\!&=&\!\!\!\!\! -\Big(\frac{q}{E_1}\Big)-
\frac{35}{4}\Big(\sin^2 \frac{3\theta}{2}\Big) \Big(\frac{q}{E_1}\Big)^2
\nonumber\\
\!\!\!\!&+&\!\!\!\!\!
\frac{35}{4}\Big(\sin^2 \frac{3\theta}{2} -\frac{35}{32}
\sin^2 3\theta \Big)\Big(\frac{q}{E_1}\Big)^3+\cdots.
\end{eqnarray}
The mean-field ground-state energy for Eq. (\ref{H}) is given per
unit cell as
\begin{eqnarray}
E^{\rm gs}_{\rm mf}=E_{\rm 1 site}^{\rm gs}(A)+E_{\rm 1 site}^{\rm gs}(B)-\frac{1}{z\lambda}{\bf q}^{A} \cdot {\bf
 q}^B, \label{E0} 
\end{eqnarray}
where $E_{\rm 1 site}^{\rm gs}(A)$ denotes the ground-state energy at
$A$-site with $q^B$ and $\theta_B$, and similar definitions for the
$B$-site.

\section{Direct products of irreducible representations of $T_d$ group} \label{direct}
In this Appendix, we list a full set of reduction tables of direct
products $\Gamma_i\otimes \Gamma_j$ in
$T_d$ point group. These are very useful for symmetry arguments
 to construct coupling constants with a
given set of operators. The lists for
$\Gamma_1\otimes \Gamma_i$ and $\Gamma_2 \otimes \Gamma_i$ are not shown, since these
are trivial. The others are as follows.

\begin{itemize}
\item $\Gamma_3\otimes\Gamma_3'=\Gamma_1\oplus\Gamma_2\oplus\Gamma_3$\\
$\Gamma_1:uu'+vv'$, \\
$\Gamma_2:uv'-vu'$, \\ 
$\Gamma_3:\big\{uu'-vv',\ -uv'-vu'\big\}$. 
\\
\item $\Gamma_3\otimes\Gamma_4=\Gamma_4\oplus\Gamma_5$\\
$\Gamma_4:\big\{(u-\sqrt{3}v)\tilde{X}, \ \ 
(u+\sqrt{3}v)\tilde{Y}, \ \ -2u \tilde{Z}
\big\}$,\\
$\Gamma_5:\big\{(\sqrt{3}u+v)\tilde{X}, \ \ 
(-\sqrt{3}u+v)\tilde{Y}, \ \ -2v \tilde{Z}
\big\}$.
\\
\item $\Gamma_3\otimes\Gamma_5=\Gamma_4\oplus\Gamma_5$\\
$\Gamma_4:\big\{(\sqrt{3}u-v){X}, \ \ 
(-\sqrt{3}u+v){Y}, \ \ -2v {Z}
\big\}$,\\
$\Gamma_5:\big\{(u-\sqrt{3}v){X}, \ \ 
(u+\sqrt{3}v){Y}, \ \ -2u {Z}
\big\}$.
\\
\item $\Gamma_4\otimes \Gamma_4^{\prime} =\Gamma_1\oplus
      \Gamma_3 \oplus \Gamma_4 \oplus \Gamma_5$\\
      $\Gamma_1:
\tilde{X}\tilde{X}'+\tilde{Y}\tilde{Y}'+ \tilde{Z}\tilde{Z}'$,\\
      $\Gamma_3:\big\{
      2\tilde{Z}\tilde{Z}'-\tilde{X}\tilde{X}'-\tilde{Y}\tilde{Y}',
       \sqrt{3}(\tilde{X}\tilde{X}'-\tilde{Y}\tilde{Y}')
\big\}$,\\
      $\Gamma_4:\big\{
\tilde{Y}\tilde{Z}'-\tilde{Z}\tilde{Y}',
\tilde{Z}\tilde{X}'-\tilde{X}\tilde{Z}',
\tilde{X}\tilde{Y}'-\tilde{Y}\tilde{X}'
\big\}$,\\
      $\Gamma_5:\big\{
\tilde{Y}\tilde{Z}'+\tilde{Z}\tilde{Y}',
\tilde{Z}\tilde{X}'+\tilde{X}\tilde{Z}',
\tilde{X}\tilde{Y}'+\tilde{Y}\tilde{X}'
\big\}$.
\\
\item $\Gamma_4\otimes \Gamma_5 =\Gamma_2\oplus
      \Gamma_3 \oplus \Gamma_4 \oplus \Gamma_5$\\
      $\Gamma_2: \tilde{X}X+\tilde{Y}Y+\tilde{Z}Z$,\\
      $\Gamma_3:\big\{
       -\sqrt{3}(\tilde{X}{X}-\tilde{Y}{Y}),
       2\tilde{Z}{Z}-\tilde{X}{X}-\tilde{Y}{Y}
\big\}$,\\
      $\Gamma_4:\big\{
\tilde{Y}{Z}+\tilde{Z}{Y},
\tilde{Z}{X}+\tilde{X}{Z},
\tilde{X}{Y}+\tilde{Y}{X}
\big\}$,\\
      $\Gamma_5:\big\{
\tilde{Y}{Z}-\tilde{Z}{Y},
\tilde{Z}{X}-\tilde{X}{Z},
\tilde{X}{Y}-\tilde{Y}{X}
\big\}$.
\\
\item $\Gamma_5\otimes \Gamma_5^{\prime}=\Gamma_1\oplus \Gamma_3 \oplus \Gamma_4
      \oplus \Gamma_5$\\
$\Gamma_1: XX'+YY'+ZZ'$,\\
$\Gamma_3: \big\{2ZZ'-XX'-YY', \sqrt{3}(XX'-YY')\big\}$,    \\
$\Gamma_4:\big\{
{Y}{Z}'-{Z}{Y}',
{Z}{X}'-{X}{Z}',
{X}{Y}'-{Y}{X}'
\big\}$,\\
      $\Gamma_5:\big\{
      {Y}{Z}'+{Z}{Y}',
{Z}{X}'+{X}{Z}',
{X}{Y}'+{Y}{X}'
\big\}$.
\end{itemize}
Finally, we show a cubic invariant in the triple product $\Gamma_3\otimes \Gamma_3^{\prime}
\otimes\Gamma''_3$,
which is directly related to Eq. (\ref{classicalmap}):
\begin{itemize}
%
\item $\Gamma_3\otimes \Gamma_3^{\prime} \otimes\Gamma''_3=\Gamma_1\oplus
      \Gamma_2 \oplus 3\Gamma_3\\
      \Gamma_1: (uu'-vv')u''-(uv'+vu')v''$,\\
 \ \ \ $\to u(u^2-3v^2)$ \ \  for $\Gamma_3 =\Gamma_3^{\prime}=\Gamma''_3$.

\end{itemize}

\section{Landau free energy for quadrupole moment}\label{Landau}

To study quadrupole anisotropy, we will in this Appendix 
calculate the corresponding Landau free energy $F(\mathbf{Q})$ 
for a single site at magnetic field $\mathbf{H}=\mathbf{0}$, 
starting from the microscopic model.  
We first consider the case of finite temperature, and 
secondly study the zero-temperature case, which needs 
a special care.  

We will calculate the quadrupole Landau free energy $F(\mathbf{Q})$ 
for a single site for a given temperature $T=\beta^{-1}$, 
with starting from the microscopic CEF 
Hamiltonian Eq. (\ref{eqHcefQ}).  
To this end, we include to the Hamiltonian a coupling 
to conjugate field $\mathbf{h}$=
$(h_z , h_x )$=$h ( \cos \theta , \sin \theta )$ and 
obtain its related free energy 
 $\tilde{F}(\mathbf{h})$:
\begin{equation}
\tilde{F}(\mathbf{h}) = 
-\frac{1}{\beta} \log \tilde{Z} (\mathbf{h}) 
=  
-\frac{1}{\beta} \log \mbox{Tr } \, 
e^{-\beta ( H_{\rm CEF} - \mathbf{h} \cdot \hat{\mathbf{Q}})}. 
\end{equation} 
The unperturbed Hamiltonian is a 9$\times$9 matrix, and 
consists of the $\Gamma_3$ ground-state doublet and excited states, 
$\Gamma_1$ singlet and two triplets, $\Gamma_4$ and $\Gamma_5$.  
With setting the ground-state energy zero $E_3=0$, 
the energies of these excited multiplets are denoted as 
$E_1$, $E_4$, and $E_5$, respectively.  

Using a standard technique, we first expand $\tilde{Z}(\mathbf{h})$ 
in $h$,
\begin{equation} 
\tilde{Z} (\mathbf{h}) = Z_0 + d_2 h^2 + d_3 h^3 + d_4 h^4 + \cdots . 
\end{equation}
Here, the unperturbed partition function is 
$Z_0 = 2 + e^{-\beta E_1} + 3 (e^{-\beta E_4} + e^{-\beta E_5})$.  
The first-order term vanishes, and this means that 
the thermal average of moment vanishes 
at any finite temperature, unless the inter-site interactions 
are switched on.  
It will turn out that the expansion up to the fourth order 
is sufficient for studying quadrupole anisotropy.  
Since the operators $\hat{\mathbf{Q}}$ are block-diagonal 
in the CEF bases in Appendix A, the expansion is easy but 
the results are not so simple.  
The second- and the fourth-order terms are isotropic with respect to 
the field direction $\theta$, while the third-order term depends 
 as $ \cos 3\theta$ and we denote $d_3=\bar{d}_3\cos 3\theta$. The
 explicit forms of $\{d_n\}$ are given as
\begin{eqnarray}
\frac{d_2}{\beta^2}
\!\!\!\!\!\!&=&\!\!\!\!\!\!  1+ \frac{35}{4}  \frac{ 1 -
e^{-\beta E_1} }{\beta E_1}
+  \frac{3}{16} 
( 7^2 e^{- \beta E_4 } + 2^2 e^{- \beta E_5 } ) 
\nonumber\\
\!\!\!\!\!\!&&\!\!\!\!\!\!+\frac{63}{32}  
\frac{e^{-\beta E_5} - e^{-\beta E_4} }{\beta(E_4 - E_5) },\\
\frac{\bar{d}_3}{\beta^3}
\!\!\!\!\!\!&=&\!\!\!\!\!\! -\frac{35}{4} 
\left[	\frac{1}{\beta E_1} 
- \frac{ 1 - e^{-\beta E_1} }{\beta^2E_1^2} 
\right] 
 \ \ \ \nonumber
\\
\!\!\!\!\!\!&&\!\!\!\!\!\!
-\frac{1}{2^7}\Bigg[ 
2  ( 7^3 e^{- \beta E_4 } - 8 e^{- \beta E_5 } ) 
+ 63  
\frac{7 e^{-\beta E_4} + 2 e^{-\beta E_5} }{\beta(E_4 - E_5) } \nonumber\\
\!\!\!\!\!\!&&
-  3^4 \cdot 7 
\frac{e^{-\beta E_5} - e^{-\beta E_4} }{\beta^2(E_4 - E_5 )^2} \Bigg],\\
\frac{d_4}{\beta^4}
\!\!\!\!\!\!&=&\!\!\!\!\!\!\frac{1}{2^5 \cdot 3} 
\Bigg[ 8  + 420 \frac{1}{\beta E_1}  
+105 \frac{27 + 35 e^{-\beta E_1}}{\beta^2 E_1^2} \nonumber\\
\!\!\!\!\!\!&&\!\!\!\!\!\!
- 6510 \frac{1 - e^{-\beta E_1}}{\beta^3E_1^3}  
\Bigg]
+ \frac{3}{2^{10}} \Bigg[
 ( 7^4 e^{- \beta E_4 } + 2^4 e^{- \beta E_5 } ) 
\nonumber\\
\!\!\!\!\!\!&&\!\!\!\!\!\!
+  21  
\frac{7^2 e^{-\beta E_4} - 4 e^{-\beta E_5} }{\beta(E_4 - E_5) } 
- \frac{3^3 \cdot 7}{4} 
\frac{7^2 e^{-\beta E_4} + 9 e^{-\beta E_5} }{\beta^2(E_4 - E_5 )^2} 
\nonumber\\
\!\!\!\!\!\!&&\!\!\!\!\!\!
+  \frac{9 \cdot 7 \cdot 29}{2} 
\frac{e^{-\beta E_5} - e^{-\beta E_4} }{\beta^2(E_4 - E_5 )^2} 
\Bigg].
\end{eqnarray}

Converting the moments $\{d_n\}$ into cumulants, the free energy is obtained 
as a series of $h$, 
\begin{equation} 
\tilde{F} ( \mathbf{h} ) 
\sim F_0 - {\textstyle \frac12} \chi h^2 
- {\textstyle \frac13} \kappa_3 h_z ( h_z^2 - 3 h_x^2 ) 
+ {\textstyle \frac14} \kappa_4 h^4, 
\end{equation}
with $F_0 = -\beta^{-1} \log Z_0$ and the coefficients are 
\begin{eqnarray} 
&&\chi = \frac{2}{\beta} \, \frac{d_2}{Z_0} 
, \ \ \ 
\kappa_3 = \frac{3}{\beta} \, \frac{\bar{d}_3}{Z_0} 
,
\nonumber\\
&&\kappa_4 = \frac{4}{\beta} \, 
\left[ 
\frac12 \left( \frac{d_2}{Z_0} \right)^2 \!\!
- \frac{d_4}{Z_0} 
\right]
= 
\frac{\beta}{2} \chi^2 - \frac{4}{\beta} \, \frac{d_4}{Z_0}.\ \ \ \ \ \
\ \ \  
\end{eqnarray}

The thermal average of the quadrupole moment is calculated as 
$\mathbf{Q}$=$(Q_z , Q_x )$=$Q (\cos \varphi , \sin \varphi )$ 
$=$$- \partial \tilde{F} (\mathbf{h} ) / \partial \mathbf{h}$ 
and the result is 
\begin{equation} 
\mathbf{Q}  \sim \chi \mathbf{h} 
+ \kappa_3 h^2 (\cos 2 \theta , -\sin 2 \theta ) - \kappa_4 h^2 \mathbf{h}. 
\end{equation}
This shows that the second-order coefficient $\chi$ 
is the linear susceptibility of quadrupole and isotropic, 
and also that the third-order contribution 
tilts the moment away from the field direction.  

The next step is to invert the relation 
and obtain for a given $\mathbf{Q}$ its corresponding $\mathbf{h}({\bf Q})$. 
This is easily done in the polar representation 
and the result is
\begin{eqnarray}
\hspace{-1cm}
h \!\!\!\!\!\!&\sim& \!\!\!\!\!\!
\frac{Q}{\chi}  
-  \frac{\kappa_3 Q^2}{\chi^3}  \cos 3 \varphi + 
\left[ \frac{\kappa_3^2}{\chi^2} 
( 1 + \frac52 \sin^2 3 \varphi ) 
+ \frac{\kappa_4}{\chi} \right] \frac{Q^3}{\chi^3} ,\nonumber\\
&& \\
\theta \!\!\!\!\!\! &\sim& \!\!\!\!\!\!
\varphi + \frac{\kappa_3 Q}{\chi^2}  \sin 3 \varphi. 
\end{eqnarray}

Now, combining all these, we can obtain the Landau 
free energy of quadrupole moment. 
We perform the Legendre transformation, 
$F(\mathbf{Q}) = \tilde{F}(\mathbf{h}(\mathbf{Q})) $%
+$ \mathbf{Q} \cdot \mathbf{h}(\mathbf{Q})$ 
and obtain  
\begin{eqnarray} 
F(\mathbf{Q}) \!\!\!\!\!&\sim& \!\!\!\!\!\!
F_0 + \frac{1}{2\chi} Q^2 
- \frac{\kappa_3}{3 \chi^3} Q_z (Q_z^2 - 3 Q_x^2 ) 
\nonumber\\
\!\!\!\!\!\!&&\!\!\!\!\!\! +\frac{1}{3 \chi^5} 
\left( 2\kappa_3^2 + \kappa_4 \chi \right) Q^4. \label{F}
\end{eqnarray}
As predicted by the symmetry argument, anisotropy 
is due to the third-order term.  

With approaching zero temperature, the linear susceptibility 
diverges as $\chi \propto \beta = 1/T$.  
Therefore, all the coefficients in the expansion above vanish, 
since $\kappa_3 \propto \beta$ and $\kappa_4 \propto \beta^3$.
This indicates that the expansion at zero temperature is not regular
around ${\bf Q}={\bf 0}$. In the $\Gamma_3$ ground-state doublet, the modulus
of quadrupole moment is $Q=1$. Therefore, at zero temperature, we need
expansion starting from $Q=1$ not 0.

At zero temperature, it is sufficient to consider the ground states and
the excited $\Gamma_1$ singlet. This is because $\Gamma_1$ is the only
state that is connected to the ground state by matrix elements of $\hat{\bf Q}$.
Solving the eigenvalue 
equation of the 3$\times$3 Hamiltonian matrix,  
we obtain the exact ground state energy 
and its expansion in $h$ reads 
\begin{equation} 
E_0 (\mathbf{h}) \sim 
- h - \frac{35}{8 E_1} h^2 
\left[ 1 + \frac{h_z ( h_z^2 - 3 h_x^2 )}{h^3} \right] .  
\end{equation}
This result is different from the expansion of the free energy 
in two points. 
First, $E_0$ has the term of order $h^1$. 
This implies the presence of spontaneous moment 
when the field $\mathbf{h}$ is switched off. 
Secondly, the anisotropy 
appears in the order $h^2$ instead of $h^3$, 
although the dependence on field direction 
is common with the finite-temperature case. 

The quadrupole moment is again obtained by 
$\mathbf{Q} = - \partial E_0 (\mathbf{h}) / \partial \mathbf{h}$ 
and the result in the polar representation is 
\begin{eqnarray}
&&\hspace{-1cm}
Q \sim 1 + \frac{35}{2E_1  } \, h \, \cos^2 \frac{3\theta}{2} 
+ \frac{105^2}{2^7 E_1 ^2} \, h^2 \,\sin^2 3 \theta \, , 
\\
&&\hspace{-1cm}
\varphi \sim \theta - \frac{105}{8E_1  } \, h \, \sin 3 \theta . 
\end{eqnarray}
As we noted above, the quadrupole moment deviates  
from $Q=1$ not 0 upon applying field at zero temperature. 
One should note that the correction terms 
vanish, when the field angle is $\theta = \frac23 \pi \times$(integer). 
This is also the case in the all orders in $h$. 
This is because the ground state does not change at all 
in this case upon increasing $h$.  

The next step is the inversion of the relation $\mathbf{Q} (\mathbf{h})$ 
and we get the field strength and angle as 
\begin{eqnarray}
\frac{h}{E_1 } \!\!\!\!\!\!  &\sim& \!\!\!\!\!\! 
\frac{Q-1}{\frac{35}{2} \cos ^2 \frac32 \varphi} 
+\frac{945}{16} 
\left( \frac{Q-1}{\frac{35}{2} \cos ^2 \frac32 \varphi} \right)^2 
\sin^2 \frac{3\varphi}{2}, \ \ \ \ \ \ 
\\
\theta \!\!\!\!\!\! &\sim& \!\!\!\!\!\! \varphi + \frac32 (Q-1) \tan \frac{3\varphi}{2}  . 
\end{eqnarray}
Note that the small parameter in expansions  
is $(Q-1)/(\frac{35}{2}\cos^2 \frac32 \varphi)$. 

Following the same procedure as for the Landau free 
energy at finite temperature, 
we obtain the zero-temperature energy of the quadrupole moment.  
The result is 
\begin{eqnarray}
\frac{E ( \mathbf{Q} ) }{E_1 }
\!\!\!\!\!\!&\sim&  \!\!\!\!\!\!
\frac{1}{35} \frac{(Q-1)^2}{\cos ^2 {\textstyle \frac32} \varphi} \nonumber\\
\!\!\!\!\!\!&-&\!\!\!\!\!\! \frac{1}{70^2}
\frac{(Q-1)^3}{\cos ^4 {\textstyle \frac32} \varphi} 
\Bigl( 2547 - 2555 \cos^2 \frac{3\varphi}{2} \Bigr).\ \ \ \ \ \ 
\label{eq:result-EQ}
\end{eqnarray}
The zero-temperature result 
has two essential differences.  
First, the anisotropy starts to appear in the second order in $(Q-1)$, 
and secondly, the angle dependence is not a single harmonic like 
$\cos 3 \varphi$.  
These two are evidences of the fact that  
the anisotropy at zero temperature cannot be 
represented by an analytic form in $\mathbf{Q}$, 
even if only small deviations are concerned.

\section{Mean-field theory for susceptibility of 
a two-sublattice system}\label{susMF}

We explain in this Appendix a method of 
calculating susceptibility in a system with 
the two sublattices $A$ and $B$. 
The input of the method is local susceptibility 
at one site of each sublattice, $\chi^s_{\rm loc}$ ($s$=$A$ or $B$).  
We prepare its exact value by either an analytic or numerical method. 
This is easy since the local Hilbert space is small, 
nine dimensions in our case.  
Let us employ a mean-field theory and 
derive a formula that gives the susceptibility 
of the entire system from the local susceptibilities. 
To be specific, we will consider the response of 
quadrupole moment $\mathbf{Q}^s$ with respect to its 
conjugate field $\mathbf{h}^s$.  

In the mean-field theory, quadrupole moment at 
each site feels an effective 
field that consists of the molecular fields contributed 
by its neighbor sites and an external conjugate field. 
This is represented by the following mean-field 
Hamiltonian
\begin{eqnarray}
H^s_{\rm MF}=H_{\rm loc}^s-({\bf h}^s -g\langle {\bf Q}^{\bar{s}}\rangle)
 \cdot {\bf Q}^s=H_0^s-{\bf h}^s\cdot {\bf Q}^s.
\end{eqnarray}
Here, the intersite coupling is $g=z\lambda$ and $s=A$ or $B$ with
$\bar{A}=B$ and vice versa.
The on-site part $H_{\mathrm{loc}}^s$ is the CEF Hamiltonian 
$H_{\mathrm{CEF}}$ plus 
the Zeeman coupling to magnetic field if it is applied.  
It does not matter if the magnetic field depends on the sublattices or not, 
and the following results hold. 
$H_0^s$ is the unperturbed part, when concerned is 
the linear response to \textit{external fields} $\mathbf{h}^{A,B}$, 
and this already includes the contribution of the molecular fields 
in ordered phases.  

Now, let us examine the linear response of $\mathbf{Q}$ at each 
sublattice.  
There are two points. 
The first is that the susceptibility is a 2$\times$2 matrix 
in $\mathbf{Q}$ space and also a 2$\times$2 matrix 
in the sublattice space
\begin{eqnarray}
\delta Q_{\mu}^s = \sum_{s,\mu'}{\chi}^{ss'}_{\mu\mu'} h^{s'}_{\mu'}
 \quad \mu,\mu'\in \{x,z\}.
\end{eqnarray}
Here, $\delta Q_{\mu}^s$ denote the induced moments  
due to the external fields, and 
we will calculate ${\chi}^{ss'}_{\mu\mu'}$.  
The second point is important and the core idea of the mean-field theory: 
the induced moments modifies the molecular fields, and 
this can be represented by the renormalization of the external fields. 
Since we know exactly the local response at each sublattice, 
they constitute self-consistency conditions
\begin{equation} 
\delta {\bf Q}^s = \underline{\chi}_{{\rm loc}}^{s} 
({\bf h}^s - g \delta {\bf Q}^{\bar{s}}). \label{eq-F:self-consistent}
\end{equation}
Here, under-bar denotes a 2$\times$2 matrix, and 
the orbital-space degrees of freedom are thus represented 
by vector and matrix for simplicity.  

It is straightforward to solve these and we obtain
\begin{eqnarray}
\underline{\chi}^{ss} &=& 
(\underline{1}-g^2\underline{\chi}_{\rm
 loc}^s\underline{\chi}_{\rm loc}^{\bar{s}})^{-1}\underline{\chi}_{\rm loc}^s,
\\
\underline{\chi}^{s\bar{s}}&=&
-g(\underline{1}-g^2\underline{\chi}_{\rm
 loc}^s\underline{\chi}_{\rm loc}^{\bar{s}})^{-1}\underline{\chi}_{\rm loc}^s
\underline{\chi}_{\rm loc}^{\bar{s}}.
%
\end{eqnarray}
The uniform and staggered susceptibilities are given as
\begin{eqnarray}
\underline{\chi}^{\pm}\!\!\!\!&=&\!\!\!\!
\underline{\chi}^{AA}+\underline{\chi}^{BB}
\pm (\underline{\chi}^{AB}+ \underline{\chi}^{BA})
,\\
\!\!\!\!&=&\!\!\!\!\sum_{s=A,B}
(\underline{1}-g^2\underline{\chi}_{\rm loc}^s\underline{\chi}_{\rm
loc}^{\bar{s}})^{-1}\underline{\chi}_{\rm loc}^s
(\underline{1}\mp g\underline{\chi}_{\rm loc}^{\bar{s}}).\ \ \ \ \ \ \ \label{finalchiMF}
\end{eqnarray}
If a staggered order is present, it induces the cross responses
\begin{eqnarray}
\underline{\chi}^{+-},\underline{\chi}^{-+}
=\underline{\chi}^{AA}
-\underline{\chi}^{BB}
\mp(
\underline{\chi}^{AB}
-\underline{\chi}^{BA}).
\end{eqnarray}
Here, $\underline{\chi}^{+-}$ and $\underline{\chi}^{-+}$ take
$-$ and $+$ sign, respectively. Beware that these are not symmetric in
the internal ${\bf Q}$ space, and the correct symmetry is 
$(\underline{\chi}^{+-})_{\mu\mu'}=
(\underline{\chi}^{-+})_{\mu'\mu}$.
 


\section{Landau theory of antiferro quadrupole order}\label{VariLandau}

In this Appendix, we study the Landau free energy 
for the antiferro quadrupole order and find 
its minimum point.  
As discussed in details in Appendix \ref{Landau}, 
the single-site free energy is given by Eq. (\ref{F}). 
Adding the inter-site interaction, we set up 
a minimal form of the total free energy 
\begin{eqnarray}
F_{\mathrm{tot}} &=& \sum_{s=A,B} 
\Bigg[ \frac{1}{2} a |\mathbf{Q}^s|^2 
- \frac{1}{3}  \gamma 
 Q_{z}^s ( {Q_{z}^s}^2 - 3{Q_{x}^s}^2 ) \nonumber\\
&&+ \frac{1}{4}  b  |\mathbf{Q}^s|^4 \Bigg] + g \mathbf{Q}^A \cdot  \mathbf{Q}^B . \label{eq-G:F}
\end{eqnarray}
Here, $a, b, \gamma$, and $g$ are all positive phenomenological constants 
and we will consider the transition with the change of $a$.  
This free energy (\ref{eq-G:F}) is appropriate for $\mathbf{H}=\mathbf{0}$ 
and also for $\mathbf{H} \parallel [111]$, 
where the magnetic field does not directly couple with $\mathbf{Q}$.  
Note that $F_{\mathrm{tot}}$ is invariant upon simultaneous 
rotation of $\mathbf{Q}^{A,B}$ by the angle $\pm \frac23 \pi$, 
but not invariant for inversion  
$\mathbf{Q}^{A,B} \rightarrow - \mathbf{Q}^{A,B}$ 
due to the $\gamma$ term.  

For simplicity, we focus on the the case of 
$|\mathbf{Q}^A| = | \mathbf{Q}^B|$. 
This corresponds to the mean-field solutions 
for $\mathbf{H}=\mathbf{0}$ and $\mathbf{H} \parallel [111]$. 
Parameterizing $\mathbf{Q}^s = (Q_z^s, Q_x^s ) 
= q ( \cos \theta_s , \sin \theta_s )$, 
the free energy is now a function of 
the two angles and the common amplitude 
\begin{equation}
F_{\mathrm{tot}} = 
a q^2 + {\textstyle \frac12} b q^4 + gq^2 R (\theta_A, \theta_B ) . 
\label{eq-G:F2}
\end{equation}
Here, the new factor is 
\begin{equation}
R ( \theta_A , \theta_B )  \equiv 
\cos ( \theta_A - \theta_B ) 
- \frac{\gamma}{3g} q \, ( \cos 3 \theta_A + \cos 3 \theta_B ) .  
\end{equation}
We minimize $F_{\mathrm{tot}}$ in two steps;
we first minimize $R$ for a given $q$ and 
secondly minimize the whole with respect to $q$.  
$R$ is minimized for a nearly antiparallel 
configuration of the two quadrupole moments; 
\begin{equation} 
\theta_{A,B} = \pm \Bigg(
\frac{\pi}{2} + \frac{\gamma}{2g} q\Bigg)
, \ \ \ 
R_{\mathrm{min}} \sim 
-1 
-\frac{\gamma^2}{2g^2} \, q^2 . 
\label{eq-G:Rmin} 
\end{equation}
There are two other solutions, but they agree 
to the above one by symmetry operation of $\pm 2\pi/3$ rotation.  

Inserting this result to Eq.~(\ref{eq-G:F2}), 
we find the function to be minimized is a standard $\phi^4$-model 
with modified coefficients
\begin{equation} 
F_{\mathrm{tot}} \sim 
\tilde{a} q^2 
+ {\textstyle \frac12} \tilde{b} q^4 , 
\hspace{0.4cm}
\tilde{a} \equiv a - g , \ \mbox{and }
\tilde{b} \equiv b - \gamma^2 /g  \, . 
\end{equation}
Here, $\tilde{b} > 0$ 
if the anisotropy is not so strong, and 
we first consider this case. 
Then, the transition is continuous. 
Finite moments start to appear when $a$ becomes 
smaller than the critical value $a_c = g$, 
and their amplitude is 
the mean-field result of the $\phi^4$-model,  
$q \sim (|\tilde{a}|/\tilde{b})^{1/2}$.  
Including higher-order corrections in the calculations 
above, we obtain the result up to the next order 
\begin{equation} 
q \sim
\sqrt{\frac{|\tilde{a}|}{\tilde{b}}}\,  
\left( 1 - \frac{\gamma^4}{2 \tilde{b}^2 g^3} | \tilde{a}| \right).  \label{qvalue}
\end{equation}
In the case of this continuous transition, 
the angle between $\mathbf{Q}^{A,B}$ starts from 
$\pi$ and gradually deforms with the growth 
of their amplitude as shown by Eq.~(\ref{eq-G:Rmin}).  

When the anisotropy is strong $\tilde{b}<0$, we need to 
include the next-order term in $q$,
\begin{equation}
F_{\mathrm{tot}} \sim 
\tilde{a} q^2 
+ \frac12 \tilde{b} q^4 
+ \frac{\gamma^2}{3g} q^6 . 
\end{equation} 
This is the $\phi^6$-model and describes a first-order 
transition upon varying $a$, when $\tilde{b}<0$.  
This transition takes place at 
\begin{equation} 
a_{\#} \sim 
g \left( 1 + \frac{3 \tilde{b}^2 g^2}{16 \gamma^4 } \right) 
\ > \ a_c , 
\end{equation}
and there the order parameter jumps to 
\begin{equation}
\bar{q}_\# \sim 
\sqrt{ \frac{3|\tilde{b}| g^3}{4\gamma^4} } . 
\label{eq-G:q1st}
\end{equation} 
Note that in the case that 
the moments $\mathbf{Q}^{A,B}$ are not completely 
antiparallel at the transition point, 
\begin{equation} 
\Delta \theta \equiv 
\theta_A - \theta_B - \pi 
\sim 
\sqrt{ \frac{3|\tilde{b}| g}{4\gamma^2} } . 
\end{equation}
Even when the anisotropy is very strong, 
the result (\ref{eq-G:q1st}) is still valid, 
but the angle deformation approaches 
$\Delta \theta \rightarrow \frac13 \pi$.  

\section{How to make invariant form} \label{how}
In this Appendix, we will discuss a general method to construct interactions that are
invariant under a given point group. Although the following discussion is
essentially the same as that done by Sakai {\it et al},\cite{Sakai} we
show some details of the calculations.

%
%

Let us consider site $i$ and its nearest neighbor sites $j=1,2,3$, and 4 in 
a diamond lattice structure. The relative positions of the nearest
neighbors ($j=1,2,3$, and 4) are 
$(111)$/4, $(\bar{1}\bar{1}1)$/4, $(1\bar{1}\bar{1})$/4, and
$(\bar{1}1\bar{1})$/4 for $j=1,2,3,4$, respectively, where we set the
lattice constant to unity. See, Fig. \ref{fig-xperp}. 
A general form of
interactions between the nearest-neighbors is 
\begin{eqnarray} 
V_i=\sum_{\mu\nu}\sum_{j} \lambda_{j}^{\mu\nu}A_i^{\mu}B_j^{\nu}. \label{eqV}
\end{eqnarray}
Here, $\lambda_{j}^{\mu\nu}$ is the coupling constant and $A^{\mu}$($B^{\nu}$)
transforms as one of the irreducible representations $\mu(\nu)$ for
local $T_d$ point group without changing the site index. We represent
the local point group operation by $R_{loc}$ and also symmetry operation
around the $i$ site on the site index $j$ by $R_{site}$. 

Now, we change the representations $\mu$ and $\nu$, which are not
irreducible with respect to the index of irreducible representation for $A\otimes
B$. We denote this index as $l$ and Eq. 
 (\ref{eqV}) is now given as
\begin{eqnarray} 
V_i=\sum_{j=1}^4 \sum_{l}\lambda_{j}^{l}[A_i\otimes B_j]_l\equiv \sum_{j=1}^4 \sum_l\lambda_{j}^{l}C_{ijl}. \label{eqV3}
\end{eqnarray}
This is further symmetrized by taking a following representation instead
of $j$:
\begin{eqnarray} 
V_i\!\!\!\!\!\!&=&\!\!\!\!\!\!\sum_l\lambda_{s}^{l}C_{il}^s
+\sum_l\vec{\lambda}^{l}\cdot \vec{C}_{il},\\ \label{eqV4}
\lambda_{s}^l\!\!\!\!\!\!&=&\!\!\!\!\!\!\frac{1}{2}\sum_{j=1}^4\lambda_{j}^l,\ 
\lambda_{x}^l=\frac{1}{2}(\lambda_{1}^l-\lambda_{2}^l+\lambda_{3}^l-\lambda_{4}^l),\\
\lambda_{y}^l\!\!\!\!\!\!&=&\!\!\!\!\!\!\frac{1}{2}(\lambda_{1}^l-\lambda_{2}^l-\lambda_{3}^l+\lambda_{4}^l),\ 
\lambda_{z}^l=\frac{1}{2}(\lambda_{1}^l+\lambda_{2}^l-\lambda_{3}^l-\lambda_{4}^l).\nonumber\\
\end{eqnarray}
Note that 
\begin{eqnarray}
(\vec{\lambda^l})_{u}=\sum_{j=1}^4\hat{U}_{u
 j}\lambda_j^l=\frac{\sqrt{3}}{2}\sum_{j=1}^4(\hat{r}_{ji})_{u
}\lambda_j^l,
\end{eqnarray}
where $\hat{r}_{ji}$ is the unit vector to $j$ from
$i$, 
and similar definitions for $C_{il}^s$ and
$(\vec{C}_{il})_{u}=\sum_j\hat{U}_{u j}C_{ijl}$. The 
site indices of $C_{il}^s$ transform as $\Gamma_1$ under the operations $R_{site}$, while that of
$\vec{C}_{il}$ as $\Gamma_5$ and $\lambda$'s do not transform under
symmetry operations $R_{site}$ because they are just coupling constants. Since
under the operations $R_{loc}$, different irreducible representations in
$l$ do not mix, we can analyze $l\in \Gamma_1$, $\Gamma_2$, $\Gamma_3$,
$\Gamma_4$ and $\Gamma_5$, separately. The result is very simple and there
are only two couplings : $\lambda_s^{\Gamma_1}$ and
$\vec{\lambda}^{\Gamma_5}$.
The latter is given by 
$\vec{\lambda}^{X}=\lambda^{\Gamma_5}(1,0,0)$, 
$\vec{\lambda}^{Y}=\lambda^{\Gamma_5}(0,1,0)$, 
$\vec{\lambda}^{Z}=\lambda^{\Gamma_5}(0,0,1)$,
 leading to 
\begin{eqnarray} 
V_i&=&\lambda_s^{\Gamma_1}\sum_{j=1}^4 C_{ij\Gamma_1}
+\lambda^{\Gamma_5} \sum_{j=1}^4\hat{r}_{ji}\cdot\  
\begin{pmatrix}
C_{ij X}\\
C_{ij Y}\\
C_{ij Z}\\
\end{pmatrix}. \ \ \ 
\label{eqV5}
\end{eqnarray}
 Equation (\ref{eqV5}) indicates that we need to construct
$\Gamma_1$ or $\Gamma_5$ representations by a given set of operators $A$
and $B$. Finally, since the above derivation does not include the fact
that the bond center of the diamond lattice structure is the inversion
center of the system, interactions obtained should be symmetrized with
respect to the inversion operations.
Even when the number of operators included increases, the above
discussion is still valid by regarding $C_{ijl}=[A_i\otimes A_i^{\prime}\otimes
\cdots \otimes B_j
\otimes B^{\prime}_j \otimes \cdots]_l$.


\begin{thebibliography}{99}
\bibitem{Th} We also note that CEF potential due to $T_h$ symmetry
$O_6^2-O_6^6$, which makes the $\Gamma_4$ and $\Gamma_5$ states in the
$T_d$ (and also $O_h$) symmetry hybridized,  can be represented also by the quadrupole operators as
\begin{eqnarray} 
O_6^2(i)-O_6^6(i)\propto Q_x^3(i)-3\overline{Q_x(i)Q_z^2(i)}\to -Q^3(i)\sin 3\theta(i).
\end{eqnarray}
\end{thebibliography}
\end{document}